\documentclass[reprint, twocolumn, secnumarabic,amssymb, aps, prb, showpacs]{revtex4-1}
\usepackage{amsmath}
\usepackage{graphicx}
\usepackage{xcolor}
\usepackage{pdfpages}

\begin{document}

\title{Understanding double-resonant Raman scattering in chiral carbon nanotubes: Diameter and energy dependence of the $D$ mode}%

\author{Felix Herziger}%
\email{fhz@physik.tu-berlin.de}

\author{Asmus Vierck}%
\author{Jan Laudenbach}%
\author{Janina Maultzsch}%

\affiliation{Institut f\"ur Festk\"orperphysik, Technische Universit\"at Berlin, Hardenbergstr. 36, 10623 Berlin, Germany}%

\date{\today}%

\begin{abstract}
We present a theoretical model to describe the double-resonant scattering process in arbitrary carbon nanotubes. We use this approach to investigate the defect-induced $D$ mode in CNTs and unravel the dependence of the $D$-mode frequency on the CNT diameter and on the energy of the resonant optical transition. Our approach is based on the symmetry of the hexagonal lattice and geometric considerations, hence the method is independent of the exact model that is chosen to describe the electronic band structure or the phonon dispersion. We finally clarify the diameter dependence of this Raman mode that was controversely discussed in the past and demonstrate that, depending on the experimental conditions, in general two different dependencies can be measured. We also prove that carbon nanotubes with arbitrary chiral index can exhibit a $D$ mode in their Raman spectrum, in contrast to previous symmetry-based arguments. Furthermore, we give a direct quantification of the curvature-induced phonon frequency corrections of the $D$-mode in carbon nanotubes with respect to graphite.
\end{abstract}


\maketitle%

\section{Introduction}
Carbon materials such as graphene and single-walled carbon nanotubes (CNTs) attracted much scientific interest in the past decades \cite{reich2004, RevModPhys.81.109}. Their extraordinary properties made them a perfect candidate for applications in novel electronic devices such as transistors or sensors \cite{10.1126/science.1060928, 10.1038/29954, 10.1038/nmat1967}. Here, functionalized systems are most important, since their properties can be designed and manipulated as preferred \cite{10.1002/1521-3773, 10.1126/science.1087691, 10.1038/nchem.1010}. The success of a functionalization is often monitored by investigating the increase of the Raman $D$-mode intensity \cite{10.1021/ja308969p}. This mode stems from TO phonons around the $K$ points of the graphene Brillouin zone and is activated by breaking the translational invariance, thus relieving selection rules \cite{10.1098/rsta.2004.1454}. Although the defect-related origin of the $D$ mode was known for a long time, the underlying Raman processes remained unclear \cite{10.1063/1.1674108}. Thomsen and Reich solved this longstanding question by introducing the double-resonant scattering theory for graphitic systems \cite{PhysRevLett.85.5214}, explaining the anomalous dispersion of certain Raman modes, \textit{e.g.}, the $D$ and $2D$ modes, with laser excitation energy \cite{PhysRevB.70.155403}. However, the double-resonant $D$ mode was thought to exist only in carbon nanotubes that satisfy $k_F = K$,\cite{PhysRevB.64.121407} where $k_F$ is the reciprocal vector of the singularity in the joint density of states. This relation is particularly true for armchair carbon nanotubes and other metallic tubes with $\mathcal{R}=3$, \textit{i.e.}, $(n_1-n_2)/(3n) = \text{integer}$ \cite{reich2004,PhysRevB.60.2728}. In contrast, semiconducting carbon nanotubes were predicted to not exhibit a $D$ mode with the same systematic excitation energy dependence of its frequency \cite{PhysRevB.64.121407}. However, recent experiments on semiconducting CNT samples enriched with single chiral indices demonstrated the existence of a $D$ mode in CNTs that do not satisfy the aforementioned restriction \cite{PhysRevB.87.165423}. Furthermore, there is an ongoing discussion about the dependence of the $D$-mode frequency in CNTs on the tube diameter and the transition energy. During the past decade many different explanations were postulated and controversely discussed. For instance, Souza Filho \textit{et al.} \cite{PhysRevB.65.035404} proposed an inverse diameter dependence of the $D$-mode frequencies, whereas Refs. \onlinecite{PhysRevB.64.041401} and \onlinecite{PhysRevB.67.035427} claimed a dependence proportional to the tube diameter. Thus, the systematics of these scattering process is still not fully understood and needs clarification.

In this work, we derive a universal, geometrical model that describes the dependence of the resonant phonon wave vector in the double-resonant scattering process on the tube diameter and the transition energy of the process for all tubes with arbitrary chiral indices, rejecting the previously predicted restriction of the $D$ mode to particular CNTs. Our model is based on the symmetry of the hexagonal lattice only and is thus universally valid, no matter which approximation for the electronic band structure or phonon dispersion is used. We apply this model to investigate the diameter dependence of the $D$ mode in CNTs for the resonant optical transition, which is still under controverse discussion. We will point out that, depending on the specific experimental conditions, in general two different diameter dependencies can be observed. Furthermore, we will highlight the importance of nanotube curvature effects on the phonon frequencies to explain the experimentally observed dispersion of the $D$ mode. Finally, we derive a diameter correction for the $D$-mode frequencies in carbon nanotubes with respect to the $D$-mode frequency in graphite.

\section{Simulation details}
All geometric considerations in this paper are based on the symmetry operations of the hexagonal lattice of graphene. Hence, within the framework of zone-folding, the results do not depend on the choice of the model to describe the electronic band structure or the phonon dispersion. 

The calculcations of electronic band structures and phonon dispersions in this work are based on the POLSym code in the sixth-nearest-neighbor approximation \cite{damnjanovic2005}. This package uses the modified group projector technique and includes curvature effects in the calculations of the band energies and phonon frequencies. Electronic bands and phonon dispersions are calculated for all 274 chiral tubes with diameters between 5 and 25\,\AA. Furthermore, we used an experimental graphite phonon dispersion from Ref. \onlinecite{PhysRevB.80.085423} for an alternative calculation of the $D$-mode frequencies and for comparison with the POLSym-derived frequencies. 

\begin{figure}
\includegraphics{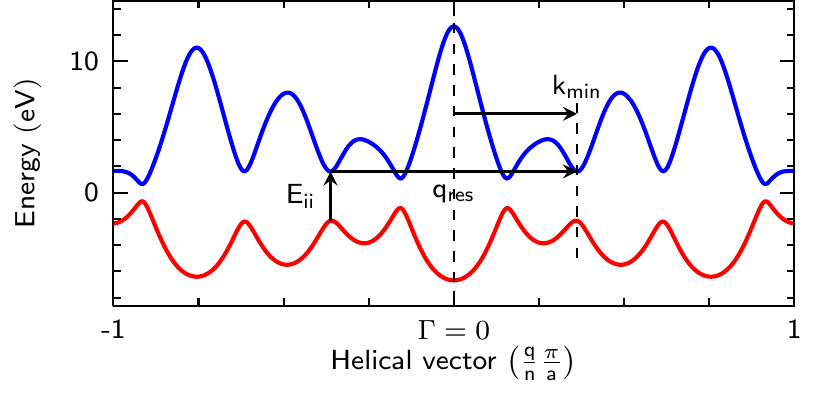}
\caption{Illustration of the double-resonant scattering process between two energetically equivalent minima in the helical band structure of a $(5,3)$-tube. The transition energy is labeled as $E_{\text{ii}}$, the position of the electronic minimum as $k_{\text{min}}$, and the resonant phonon wave vector as $q_{\text{res}}$. The translational period is denoted with $a$; $q$ reflects the subbands in linear quantum numbers. \label{fig:doubleresonance}}
\end{figure}

We model the double-resonant scattering process in carbon nanotubes by assuming that scattering occurs mostly between equivalent minima in the electronic bandstructure (compare Fig.~\ref{fig:doubleresonance}) \cite{PhysRevB.87.165423}. As demonstrated by previous works, the double-resonance process is dominated by the incoming resonance with excitonic transitions at van-Hove singularities in the electronic bandstructure \cite{PhysRevB.65.165433,10.1002/pssb.201200175}, justifying our assumption. Furthermore, we assume that the phonon frequency $\hbar\omega_{\text{phonon}} = 0$\,meV, as the resonant phonon wave-vectors in the double-resonance process depend only weakly on the energy of the involved phonons \cite{PhysRevB.84.035433}. These assumptions do not affect the general validity of our results in this work.

\section{Results and discussion}
\begin{figure}
\includegraphics{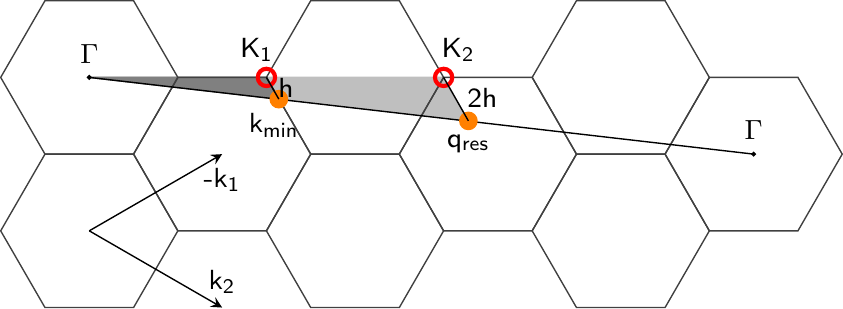}
\caption{Illustration of the dependence between the electronic transition and the resonant phonon wave-vector for a coprime tube, \textit{i.e.}, $n=1$. The black solid line denotes the helical vector of a $(3,2)$-tube, the full orange circles mark a minimum in the electronic band structure ($k_{\text{min}}$) and the corresponding resonant phonon wave-vector length ($q_{\text{res}}$), which is twice $k_{\text{min}}$. The open red circles denote the closest $K$ points to $k_{\text{min}}$ and $q_{\text{res}}$, respectively. As can be seen easily, the distance $h = K_1-k_{\text{min}}$ is always twice the distance $K_2-q_{\text{res}}$.  \label{fig:hexgrid}}
\end{figure}

As we mainly consider chiral carbon nanotubes, it is convenient to use the helical representation of the wave vector and electronic bands for the following analysis \cite{10.1088/0305-4470/33/37/308,10.1016/S0038-1098(02)00025-X}. In the zone-folding approach, the electronic band structure in helical quantum numbers is given by the cut of the helical wave vector $\mathbf{\widetilde{k}}_z$ with the electronic $\pi$ and $\pi^{*}$ bands of graphene along its path in reciprocal space. The helical wave vector is given by
\begin{equation*}
\mathbf{\widetilde{k}}_z = \frac{1}{n}\,\left(-n_2\,\mathbf{k}_1 + n_1\,\mathbf{k}_2\right) + \Delta\mathbf{k}_1(n_1,n_2) + \Delta\mathbf{k}_2(n_1,n_2),
\end{equation*}
where $n_1$ and $n_2$ are the chiral indices of the CNT and $n$ is the greatest common divisor of $(n_1,n_2)$. $\mathbf{k}_1$ and $\mathbf{k}_2$ are the unit vectors in graphene's reciprocal space (compare Fig.~\ref{fig:hexgrid}). The $\Delta\mathbf{k}_i$ describe the distance in reciprocal space of the $i$th subband from the subband containing the $\Gamma$ point of graphene and is non-zero only if $n>1$. For a more detailed derivation of the helical wave vector $\mathbf{\widetilde{k}}_z$ and an introduction to the concept of helical zone-folding we refer the reader to the Supplementary Material \footnote{See the Supplemental Material at http: which includes Refs. \onlinecite{10.1088/0305-4470/33/37/308, 10.1016/S0038-1098(02)00025-X, reich2004, 10.1016/0009-2614(93)87205-H, PhysRevLett.93.177401, PhysRevB.72.205438, PhysRevB.74.115415, PhysRevB.71.035416, 10.1016/j.carbon.2011.06.076, PhysRevB.74.184306, PhysRevB.59.8355, 10.1021/nn2044356,PhysRevB.60.2728,LaudenbachDissertation}.}. 

\begin{figure*}
\includegraphics{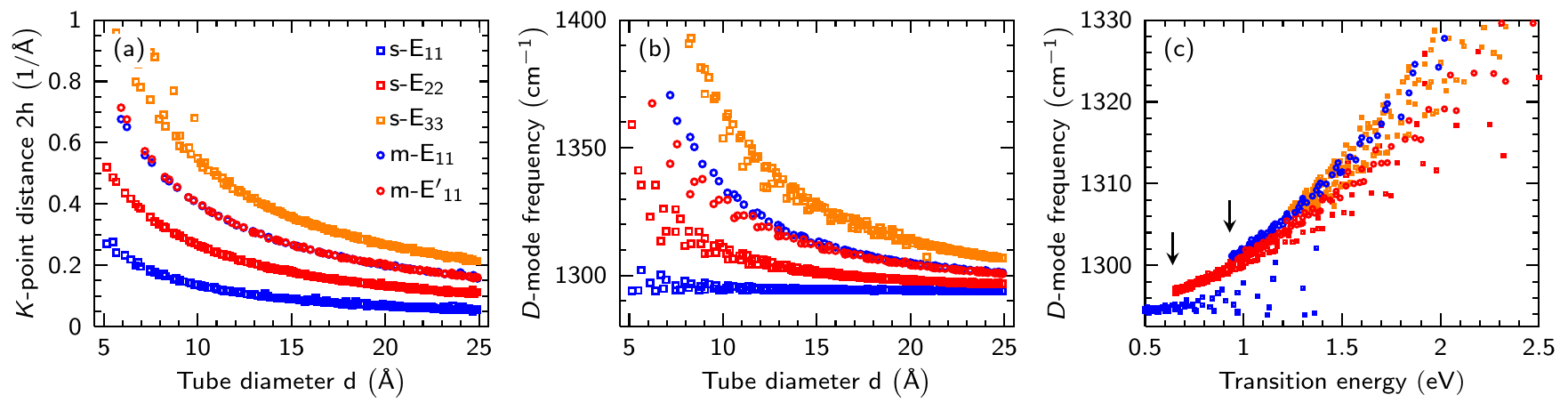}
\caption{(a) Distance between the resonant phonon wave-vector and the closest $K$ point as a function of the tube diameter for the resonantly enhanced scattering process obtained by a sixth-nearest neighbor tight-binding model for electronic band structures and phonon dispersions. (b) Calculated $D$-mode frequency $\mathsf{\omega_{D}}$ as a function of the tube diameter. (c) Calculated $D$-mode frequency as a function of the laser energy. The experimentally observed jumps between the different transitions are indicated. \label{fig:Dmodedependencies}}
\end{figure*}

The $D$ mode results from a double-resonant intervalley scattering process, including a transverse optical (TO) phonon and a defect \cite{PhysRevLett.85.5214,PhysRevB.64.121407}. In Figure \ref{fig:hexgrid} we show the systematics of the $D$-mode scattering process for a $(3,2)$-tube. Our approach is of course valid for all other tubes having more than one subband (see Supplementary Material). Without loss of generality, we assume that the minimum in the band structure of a CNT occurs at the position where the helical vector crosses a $K-M-K'$ high-symmetry line. The minimum shall have the $k$-vector $k_{\text{min}}$ (measured from the $\Gamma$ point), hence the resonant phonon wave vector in the $D$-mode scattering process has the length $q_{\text{res}} = 2\,k_{\text{min}}$. This implies the so-called $q_{\text{res}} \approx 2\,k_{\text{min}}$ rule. We label the $K$ point closest to the electronic minimum as $K_1$ and refer to its distance to $k_{\text{min}}$ as $h$. In general, the distance $h$ is a measure for the transition energy, \textit{i.e.}, a smaller value for $h$ means a lower energy and vice versa. The three points $\Gamma$, $k_{\text{min}}$, and $K_1$ form a triangle, which is indicated by the dark-gray area in Fig.~\ref{fig:hexgrid}. By similarity transformation of this triangle, we obtain the light-gray triangle formed by the points $\Gamma$, $q_{\text{res}}$, and $K_2$. It is an intrinsic property of the hexagonal lattice that twice the distance between $\Gamma$ and a $K$ point, is again a $K$ point ($K_2$). If $K_1$ is the closest $K$ point to $k_{\text{min}}$, then $K_2$ is the closest $K$ point to $q_{\text{res}}$. Since all sides of the larger triangle were doubled, the distance $K_2-q_{\text{res}}$ is now twice the distance $h$. This means that an electronic transition with a distance $h$ from $K_1$ results in a resonant phonon vector with a distance $2h$ from $K_2$ (at the same angle with respect to the closest $K$ point). Thus, $h$ and $2h$ are indicative for the energy of the transition and the $D$-mode frequency, \textit{i.e.}, a larger value $2h$ in general means a larger TO phonon frequency and vice versa. From this we can deduce two important results: First, tubes with large diameters and thus lower transition energies exhibit a systematically lower $D$-mode frequency for the resonantly enhanced transition compared to small-diameter tubes. Second, different transitions in a CNT exhibit different $D$-mode frequencies. Although these results seem trivial at a first glance, they enable a deeper understanding of the $D$ mode in CNTs.

If we now also consider trigonal warping effects in the electronic bandstructure, the electronic minimum may not be exactly on the $K-M-K'$ high-symmetry direction, but slightly shifted away. This would lead to a slightly different distance $h$ and angle between $k_{\text{min}}$ and $K_1$. Nevertheless, the distance between the resonant phonon wave vector $q_{\text{res}}$ and $K_2$ is again $2h$. Hence, this relation is independent from the exact position of the electronic minimum, which may differ depending on the model used in bandstructure calculations. Thus, our results are universally valid as long as zone-folding is an appropriate model to describe the properties of CNTs \cite{PhysRevB.70.115407,PhysRevB.65.153405,10.1088/1367-2630/6/1/017}.

Figure~\ref{fig:Dmodedependencies}\,(a) presents the calculated distance between the resonant phonon wave vector and the closest $K$ point for different optical transitions as a function of the tube diameter $d$. As can be seen, the distance $2h$ decreases with a $1/d$-dependence. Ergo, the resonantly enhanced TO phonon frequency decreases likewise. Furthermore, energetically higher transitions have a larger distance $2h$ for the same tube diameter, corresponding to higher $D$-mode frequencies. Since the distance $2h$ directly depends on the energy of the optical transition of a tube, we observe a close correspondence between the dependence shown in Fig.~\ref{fig:Dmodedependencies}\,(a) and the so-called Kataura plot \cite{10.1016/S0379-6779(98)00278-1}. By translating the distance $2h$ into a phonon frequency using the POLSym-calculated phonon dispersions, we obtain the $D$-mode frequencies shown in Fig.~\ref{fig:Dmodedependencies}\,(b). As discussed before, a larger value for $2h$ reflects a larger TO phonon frequency and vice versa, thus, the systematics of this plot again resemble the Kataura plot. We observe a decreasing $D$-mode frequency with increasing tube diameter. Furthermore, electronic transitions with a higher energy show an increased frequency.

\begin{figure}
\includegraphics{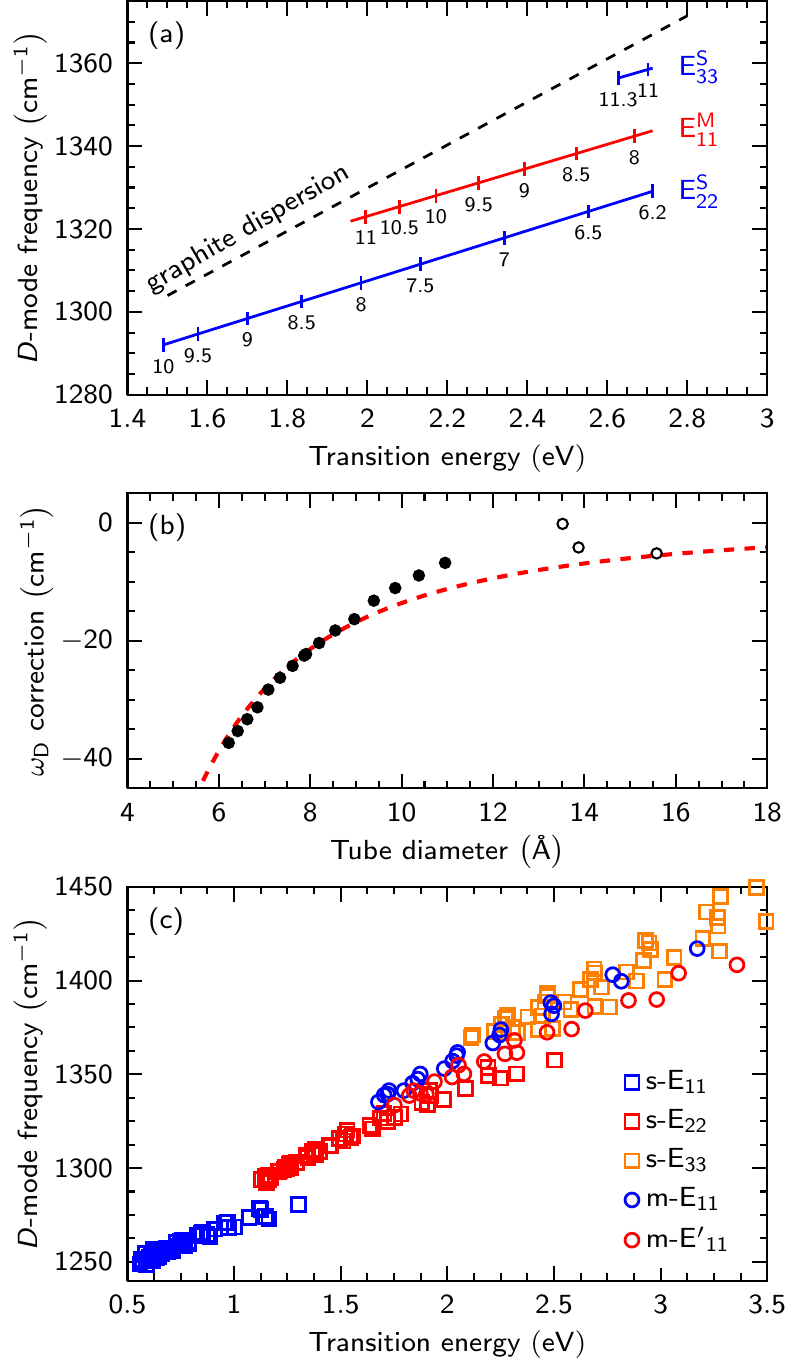}
\caption{(a) Schematized experimental data of the $D$ mode in carbon nanotubes (after Refs. \onlinecite{LaudenbachDissertation,Laudenbach2015}). The solid lines denote fits to the experimental data for each transition; the numbers at each line correspond to the diameter of the tubes (in units of \AA). The dashed line indicates the $D$-mode dispersion in graphite. (b) \textit{D}-mode frequency correction as a funtion of the tube diameter. The red, dashed line reflects a fit to the data points. Solid symbols were obtained from (a), open symbols are single-tube measurements from Refs. \onlinecite{LaudenbachDissertation,Laudenbach2015}. (c) Curvature-corrected $D$-mode dispersion from applying Eq.~\eqref{eq:correction} on the graphite dispersion. The experimentally observed discontinuity could be reproduced. \label{fig:curvature}}
\end{figure}

If we plot the $D$-mode frequency as a function of the transition energy in Fig.~\ref{fig:Dmodedependencies}\,(c), we observe an interesting feature in the $D$-mode dispersion. The dispersion of the branches is not continuous. In fact, a small jump in frequency between different electronic transitions is observed. This frequency jump has been reported previously in resonance Raman studies on enriched carbon nanotube samples \cite{LaudenbachDissertation,Laudenbach2015,10.1002/pssb.201200175}. As we discuss below, the observed jump is a direct consequence of curvature effects on the phonon frequencies and thus cannot be explained using a simple zone-folding approach. Considering a zone-folding third-nearest-neighbor tight-binding model without any curvature effects, only the helical vector would depend on $n_1$ and $n_2$, but not the electronic bands nor the phonon dispersion. Hence, two different tubes with different diameters but the same transition energy, \textit{i.e.}, same distance $h$, would exhibit the same $D$-mode frequency, \textit{i.e.}, same distance $2h$. Thus, by neglecting curvature effects, the experimentally observed frequency jump cannot be explained. However, it is well-known that phonons in carbon nanotubes show a strong dependence on the tube diameter due to curvature effects and rehybridization of $\sigma$ and $\pi$ orbitals. For instance, in semiconducting tubes, the $\Gamma$-point TO phonon frequency ($G^-$ mode) decreases with decreasing tube diameter \cite{10.1021/nn2044356}; a similar diameter dependence of the TO branch is expected around the $K$ point. Thus, two different tubes with different diameters but similar transition energies will not have the same $D$-mode frequency, \textit{i.e.}, the smaller tube has a lower frequency compared to the larger tube. Ergo, in a resonant measurement, one would observe a jump between different transition branches. In our calculations, this discontinuity is smaller than experimentally observed. Furthermore, $D$ mode from the $E_{11}^S$ transition shows nearly no dispersion with transition energy. We attribute these discrepancies to the fact that the Kohn anomaly and the region around the $K$ point in the TO phonon dispersion is not well approximated by the model for the phonon dispersion. Since the resonant phonons from the $E_{11}^S$ transition stem from a region close to the $K$ point, small deviations in the phonon dispersion directly influence the results and, in our case, lead to an overestimated $D$-mode frequency with nearly no dispersion. We expect a larger jump and a better correspondence to the experiments for calculations using a non-orthogonal tight-binding model or an \textit{ab-initio} approach. Nevertheless, the existence of a jump can be reproduced in our simulations, as seen between $E_{11}^S$ and $E_{22}^S$ in Fig.~\ref{fig:curvature}\,(c).

Next, we want to investigate the diameter distribution along a given optical transition E$_{ii}$. The energy of a transition is inversely proportional to the CNT diameter, hence, when following a $D$-mode branch while decreasing the transition energy [see Fig.~\ref{fig:Dmodedependencies}\,(c)], the diameter decreases. 

The above presented results finally harmonize the different conclusions from various previous works where a decreasing $D$-mode frequency for a decreasing tube diameter was claimed \cite{PhysRevB.64.041401,PhysRevB.67.035427} or vice versa \cite{PhysRevB.65.035404}. It is very important to point out that, depending on the experimental conditions, one can generally observe two different dispersion behaviors. By measuring the same optical transition for all tubes, \textit{e.g.}, the E$_{22}$ transition, an $\propto$1/$d$-dependence will be observed due to the mentioned diameter distribution along the E$_{22}$ branch. This would correspond to an experimental setup where the laser energy is tuned to measure every tube always at its resonance energy. By using only a single or just very few laser energies, a different behaviour is observed \cite{PhysRevB.64.041401,PhysRevB.67.035427}. This experimental condition corresponds to a vertical cut through the dispersion relation shown in Fig.~\ref{fig:Dmodedependencies}\,(c). Along such a cut, a higher $D$-mode frequency corresponds to a higher tube diameter and a dependence proportional to $d$ is observed. Both cases are distinctly different and must be separated carefully.

To finalize our results, we want to give an explicit expression for the correction of the $D$-mode frequency as a function of the tube diameter, as a correction to the $D$ mode in graphite. As pointed out before, the phonon frequencies in CNTs show a strong dependence on the tube diameter. This dependence was shown both experimentally and theoretically for the RBM \cite{PhysRevB.58.R8869,PhysRevB.73.085407,PhysRevLett.93.177401}, as well as for the G$^-$ mode \cite{10.1021/nn2044356}. However, a direct evaluation of this dependence for the $D$ mode is still lacking. Starting from the schematized experimental $D$-mode dispersion shown in Fig.~\ref{fig:curvature}\,(a) taken from Refs. \onlinecite{LaudenbachDissertation, Laudenbach2015} and the experimentally known dispersion of the $D$ mode in graphene and graphite \cite{10.1098/rsta.2004.1454,10.1038/nnano.2013.46}, we calculate the frequency difference between both dispersions and plot them as a function of the tube diameter. Since the $D$ mode in graphene and graphite does not include any curvature effects, the difference between those dispersions is a direct measurement of the diameter-dependent correction of the $D$-mode frequency. The result of this calculation is shown by filled and open circles in Fig.~\ref{fig:curvature}\,(b). As expected, the frequency correction is always negative and converges to zero for increasing tube diameter. The dashed line reflects a fit to the data derived from a classical mechanical model (see Supplementary Material), using the formula:
\begin{equation}
\Delta\omega_{D}(d) = A\,\left[\sqrt{1-\left(\frac{2.13\,\text{\AA}}{d}\right)^2} -1\right].
\label{eq:correction}
\end{equation}
Here, $d$ is the tube diameter and $A$ is a fit parameter obtained from fitting the data points in Fig.~\ref{fig:curvature}\,(b) and was determined as $A=593\,$cm$^{-1}$. 

In the following, we will apply the above result to calculated $D$-mode frequencies obtained from a phonon dispersion that does not include curvature-induced effects. We use the calculated electronic minima and resonant phonon wave vectors from our sixth-nearest-neighbor model and obtain the resonant phonon frequencies from an experimentally measured TO phonon dispersion of graphite from Ref. \onlinecite{PhysRevB.80.085423}. The CNT diameters in this calculation were chosen to fit the experimentally measured nanotubes from Refs. \onlinecite{LaudenbachDissertation, Laudenbach2015}, \textit{i.e.}, only tubes with diameters between 6\,\AA{} and 14\,\AA{} were considered. Without curvature corrections, a linear dependence between transition energy and $D$-mode frequency is observed. By applying the diameter-dependent frequency correction according to Eq.~\eqref{eq:correction}, CNTs with smaller diameter observe a larger frequency downshift than nanotubes with large diameters with respect to the graphite $D$-mode dispersion. Thus, the slope of the $D$ mode dispersion in CNTs is reduced compared to the dispersion in graphite [compare Fig.~\ref{fig:curvature}\,(a)]. As a consequence, a discontinuity opens between different transitions. In our curvature-corrected $D$-mode dispersion from Fig.~\ref{fig:curvature}\,(c), this discontinuity can be seen again most clearest between $E_{11}^S$ and $E_{22}^S$ (approx. 15\,cm$^{-1}$). For energetically higher transitions, the frequency jump decreases, \textit{e.g.}, the difference between $E_{22}^S$ and $E_{11}^M$ is approx. 9\,cm$^{-1}$. Although the calculated frequency difference is smaller than experimentally observed, the experimentally observed discontinuity is reproduced in our simulations and is shown to be a consequence of the diameter dependence of phonon frequencies in carbon nanotubes.

\section{Conclusion}
In summary, we derived a theroretical model to describe the double-resonant scattering process in arbitrary CNTs. We applied this model to describe the diameter dependence of the $D$ mode in carbon nanotubes for the resonant optical transitions. The presented approach is independent of the specific calculational model used for the electronic band structure or phonon dispersion and is therefore universally valid. We showed that, depending on the experimental conditions, in general two different diameter dependences can be observed. Furthermore, we proved that the experimentally observed discontuinity between different transition branches is due to curvature effects that alter the TO phonon dispersion at the $K$ point. Finally, we derived a quantification of the diameter-dependent frequency correction of the $D$ mode in carbon nanotubes with respect to the $D$ mode in graphite. The presented results answer the long-standing question regarding the diameter dependence of the $D$ in carbon nanotubes. Finally, we want to explicitly point out that our approach is equally valid for all other double-resonant Raman modes in CNTs and can be easily expanded to intravalley scattering processes.

\appendix
\begin{acknowledgments}
The authors thank I. Milo\v{s}evi\'{c} (University of Belgrade, Serbia) for useful discussions and kind support with the POLSym code. This work was supported by the European Research Council (ERC) under grant no. 259286 and by the DFG under grant number MA 4079/7-2. 
\end{acknowledgments}

\bibliographystyle{apsrev4-1}

\begin{thebibliography}{45}%
\makeatletter
\providecommand \@ifxundefined [1]{%
 \@ifx{#1\undefined}
}%
\providecommand \@ifnum [1]{%
 \ifnum #1\expandafter \@firstoftwo
 \else \expandafter \@secondoftwo
 \fi
}%
\providecommand \@ifx [1]{%
 \ifx #1\expandafter \@firstoftwo
 \else \expandafter \@secondoftwo
 \fi
}%
\providecommand \natexlab [1]{#1}%
\providecommand \enquote  [1]{``#1''}%
\providecommand \bibnamefont  [1]{#1}%
\providecommand \bibfnamefont [1]{#1}%
\providecommand \citenamefont [1]{#1}%
\providecommand \href@noop [0]{\@secondoftwo}%
\providecommand \href [0]{\begingroup \@sanitize@url \@href}%
\providecommand \@href[1]{\@@startlink{#1}\@@href}%
\providecommand \@@href[1]{\endgroup#1\@@endlink}%
\providecommand \@sanitize@url [0]{\catcode `\\12\catcode `\$12\catcode
  `\&12\catcode `\#12\catcode `\^12\catcode `\_12\catcode `\%12\relax}%
\providecommand \@@startlink[1]{}%
\providecommand \@@endlink[0]{}%
\providecommand \url  [0]{\begingroup\@sanitize@url \@url }%
\providecommand \@url [1]{\endgroup\@href {#1}{\urlprefix }}%
\providecommand \urlprefix  [0]{URL }%
\providecommand \Eprint [0]{\href }%
\providecommand \doibase [0]{http://dx.doi.org/}%
\providecommand \selectlanguage [0]{\@gobble}%
\providecommand \bibinfo  [0]{\@secondoftwo}%
\providecommand \bibfield  [0]{\@secondoftwo}%
\providecommand \translation [1]{[#1]}%
\providecommand \BibitemOpen [0]{}%
\providecommand \bibitemStop [0]{}%
\providecommand \bibitemNoStop [0]{.\EOS\space}%
\providecommand \EOS [0]{\spacefactor3000\relax}%
\providecommand \BibitemShut  [1]{\csname bibitem#1\endcsname}%
\let\auto@bib@innerbib\@empty
%</preamble>
\bibitem [{\citenamefont {Reich}\ \emph {et~al.}(2004)\citenamefont {Reich},
  \citenamefont {Thomsen},\ and\ \citenamefont {Maultzsch}}]{reich2004}%
  \BibitemOpen
  \bibfield  {author} {\bibinfo {author} {\bibfnamefont {S.}~\bibnamefont
  {Reich}}, \bibinfo {author} {\bibfnamefont {C.}~\bibnamefont {Thomsen}}, \
  and\ \bibinfo {author} {\bibfnamefont {J.}~\bibnamefont {Maultzsch}},\
  }\href@noop {} {\emph {\bibinfo {title} {Carbon {N}anotubes: {B}asic
  {C}oncepts and {P}hysical {P}roperties}}}\ (\bibinfo  {publisher} {Wiley-VCH,
  Weinheim},\ \bibinfo {year} {2004})\BibitemShut {NoStop}%
\bibitem [{\citenamefont {Castro~Neto}\ \emph {et~al.}(2009)\citenamefont
  {Castro~Neto}, \citenamefont {Guinea}, \citenamefont {Peres}, \citenamefont
  {Novoselov},\ and\ \citenamefont {Geim}}]{RevModPhys.81.109}%
  \BibitemOpen
  \bibfield  {author} {\bibinfo {author} {\bibfnamefont {A.~H.}\ \bibnamefont
  {Castro~Neto}}, \bibinfo {author} {\bibfnamefont {F.}~\bibnamefont {Guinea}},
  \bibinfo {author} {\bibfnamefont {N.~M.~R.}\ \bibnamefont {Peres}}, \bibinfo
  {author} {\bibfnamefont {K.~S.}\ \bibnamefont {Novoselov}}, \ and\ \bibinfo
  {author} {\bibfnamefont {A.~K.}\ \bibnamefont {Geim}},\ }\href {\doibase
  10.1103/RevModPhys.81.109} {\bibfield  {journal} {\bibinfo  {journal} {Rev.
  Mod. Phys.}\ }\textbf {\bibinfo {volume} {81}},\ \bibinfo {pages} {109}
  (\bibinfo {year} {2009})}\BibitemShut {NoStop}%
\bibitem [{\citenamefont {Baughman}\ \emph {et~al.}(2002)\citenamefont
  {Baughman}, \citenamefont {Zakhidov},\ and\ \citenamefont
  {de~Heer}}]{10.1126/science.1060928}%
  \BibitemOpen
  \bibfield  {author} {\bibinfo {author} {\bibfnamefont {R.~H.}\ \bibnamefont
  {Baughman}}, \bibinfo {author} {\bibfnamefont {A.~A.}\ \bibnamefont
  {Zakhidov}}, \ and\ \bibinfo {author} {\bibfnamefont {W.~A.}\ \bibnamefont
  {de~Heer}},\ }\href {\doibase 10.1126/science.1060928} {\bibfield  {journal}
  {\bibinfo  {journal} {Science}\ }\textbf {\bibinfo {volume} {297}},\ \bibinfo
  {pages} {787} (\bibinfo {year} {2002})}\BibitemShut {NoStop}%
\bibitem [{\citenamefont {Tans}\ \emph {et~al.}(1998)\citenamefont {Tans},
  \citenamefont {Verschueren},\ and\ \citenamefont {Dekker}}]{10.1038/29954}%
  \BibitemOpen
  \bibfield  {author} {\bibinfo {author} {\bibfnamefont {S.~J.}\ \bibnamefont
  {Tans}}, \bibinfo {author} {\bibfnamefont {A.~R.~M.}\ \bibnamefont
  {Verschueren}}, \ and\ \bibinfo {author} {\bibfnamefont {C.}~\bibnamefont
  {Dekker}},\ }\href@noop {} {\bibfield  {journal} {\bibinfo  {journal}
  {Nature}\ }\textbf {\bibinfo {volume} {393}},\ \bibinfo {pages} {49}
  (\bibinfo {year} {1998})}\BibitemShut {NoStop}%
\bibitem [{\citenamefont {Schedin}\ \emph {et~al.}(2007)\citenamefont
  {Schedin}, \citenamefont {Geim}, \citenamefont {Morozov}, \citenamefont
  {Hill}, \citenamefont {Blake}, \citenamefont {Katsnelson},\ and\
  \citenamefont {Novoselov}}]{10.1038/nmat1967}%
  \BibitemOpen
  \bibfield  {author} {\bibinfo {author} {\bibfnamefont {F.}~\bibnamefont
  {Schedin}}, \bibinfo {author} {\bibfnamefont {A.~K.}\ \bibnamefont {Geim}},
  \bibinfo {author} {\bibfnamefont {S.~V.}\ \bibnamefont {Morozov}}, \bibinfo
  {author} {\bibfnamefont {E.~W.}\ \bibnamefont {Hill}}, \bibinfo {author}
  {\bibfnamefont {P.}~\bibnamefont {Blake}}, \bibinfo {author} {\bibfnamefont
  {M.~I.}\ \bibnamefont {Katsnelson}}, \ and\ \bibinfo {author} {\bibfnamefont
  {K.~S.}\ \bibnamefont {Novoselov}},\ }\href {\doibase 10.1038/nmat1967}
  {\bibfield  {journal} {\bibinfo  {journal} {Nature Materials}\ }\textbf
  {\bibinfo {volume} {6}},\ \bibinfo {pages} {652} (\bibinfo {year}
  {2007})}\BibitemShut {NoStop}%
\bibitem [{\citenamefont {Hirsch}(2002)}]{10.1002/1521-3773}%
  \BibitemOpen
  \bibfield  {author} {\bibinfo {author} {\bibfnamefont {A.}~\bibnamefont
  {Hirsch}},\ }\href {\doibase 10.1002/1521-3773} {\bibfield  {journal}
  {\bibinfo  {journal} {Angewandte Chemie International Edition}\ }\textbf
  {\bibinfo {volume} {41}},\ \bibinfo {pages} {1853} (\bibinfo {year}
  {2002})}\BibitemShut {NoStop}%
\bibitem [{\citenamefont {Strano}\ \emph {et~al.}(2003)\citenamefont {Strano},
  \citenamefont {Dyke}, \citenamefont {Usrey}, \citenamefont {Barone},
  \citenamefont {Allen}, \citenamefont {Shan}, \citenamefont {Kittrell},
  \citenamefont {Hauge}, \citenamefont {Tour},\ and\ \citenamefont
  {Smalley}}]{10.1126/science.1087691}%
  \BibitemOpen
  \bibfield  {author} {\bibinfo {author} {\bibfnamefont {M.~S.}\ \bibnamefont
  {Strano}}, \bibinfo {author} {\bibfnamefont {C.~A.}\ \bibnamefont {Dyke}},
  \bibinfo {author} {\bibfnamefont {M.~L.}\ \bibnamefont {Usrey}}, \bibinfo
  {author} {\bibfnamefont {P.~W.}\ \bibnamefont {Barone}}, \bibinfo {author}
  {\bibfnamefont {M.~J.}\ \bibnamefont {Allen}}, \bibinfo {author}
  {\bibfnamefont {H.}~\bibnamefont {Shan}}, \bibinfo {author} {\bibfnamefont
  {C.}~\bibnamefont {Kittrell}}, \bibinfo {author} {\bibfnamefont {R.~H.}\
  \bibnamefont {Hauge}}, \bibinfo {author} {\bibfnamefont {J.~M.}\ \bibnamefont
  {Tour}}, \ and\ \bibinfo {author} {\bibfnamefont {R.~E.}\ \bibnamefont
  {Smalley}},\ }\href {\doibase 10.1126/science.1087691} {\bibfield  {journal}
  {\bibinfo  {journal} {Science}\ }\textbf {\bibinfo {volume} {301}},\ \bibinfo
  {pages} {1519} (\bibinfo {year} {2003})}\BibitemShut {NoStop}%
\bibitem [{\citenamefont {Englert}\ \emph {et~al.}(2011)\citenamefont
  {Englert}, \citenamefont {Dotzer}, \citenamefont {Yang}, \citenamefont
  {Schmid}, \citenamefont {Papp}, \citenamefont {Gottfried}, \citenamefont
  {Steinr\"uck}, \citenamefont {Spiecker}, \citenamefont {Hauke},\ and\
  \citenamefont {Hirsch}}]{10.1038/nchem.1010}%
  \BibitemOpen
  \bibfield  {author} {\bibinfo {author} {\bibfnamefont {J.~M.}\ \bibnamefont
  {Englert}}, \bibinfo {author} {\bibfnamefont {C.}~\bibnamefont {Dotzer}},
  \bibinfo {author} {\bibfnamefont {G.}~\bibnamefont {Yang}}, \bibinfo {author}
  {\bibfnamefont {M.}~\bibnamefont {Schmid}}, \bibinfo {author} {\bibfnamefont
  {C.}~\bibnamefont {Papp}}, \bibinfo {author} {\bibfnamefont {J.~M.}\
  \bibnamefont {Gottfried}}, \bibinfo {author} {\bibfnamefont {H.~P.}\
  \bibnamefont {Steinr\"uck}}, \bibinfo {author} {\bibfnamefont
  {E.}~\bibnamefont {Spiecker}}, \bibinfo {author} {\bibfnamefont
  {F.}~\bibnamefont {Hauke}}, \ and\ \bibinfo {author} {\bibfnamefont
  {A.}~\bibnamefont {Hirsch}},\ }\href {\doibase 10.1038/nchem.1010} {\bibfield
   {journal} {\bibinfo  {journal} {Nature Chemistry}\ }\textbf {\bibinfo
  {volume} {3}},\ \bibinfo {pages} {279} (\bibinfo {year} {2011})}\BibitemShut
  {NoStop}%
\bibitem [{\citenamefont {Maeda}\ \emph {et~al.}(2012)\citenamefont {Maeda},
  \citenamefont {Saito}, \citenamefont {Akamatsu}, \citenamefont {Chiba},
  \citenamefont {Ohno}, \citenamefont {Okui}, \citenamefont {Yamada},
  \citenamefont {Hasegawa}, \citenamefont {Kako},\ and\ \citenamefont
  {Akasaka}}]{10.1021/ja308969p}%
  \BibitemOpen
  \bibfield  {author} {\bibinfo {author} {\bibfnamefont {Y.}~\bibnamefont
  {Maeda}}, \bibinfo {author} {\bibfnamefont {K.}~\bibnamefont {Saito}},
  \bibinfo {author} {\bibfnamefont {N.}~\bibnamefont {Akamatsu}}, \bibinfo
  {author} {\bibfnamefont {Y.}~\bibnamefont {Chiba}}, \bibinfo {author}
  {\bibfnamefont {S.}~\bibnamefont {Ohno}}, \bibinfo {author} {\bibfnamefont
  {Y.}~\bibnamefont {Okui}}, \bibinfo {author} {\bibfnamefont {M.}~\bibnamefont
  {Yamada}}, \bibinfo {author} {\bibfnamefont {T.}~\bibnamefont {Hasegawa}},
  \bibinfo {author} {\bibfnamefont {M.}~\bibnamefont {Kako}}, \ and\ \bibinfo
  {author} {\bibfnamefont {T.}~\bibnamefont {Akasaka}},\ }\href {\doibase
  10.1021/ja308969p} {\bibfield  {journal} {\bibinfo  {journal} {Journal of the
  American Chemical Society}\ }\textbf {\bibinfo {volume} {134}},\ \bibinfo
  {pages} {18101} (\bibinfo {year} {2012})}\BibitemShut {NoStop}%
\bibitem [{\citenamefont {Reich}\ and\ \citenamefont
  {Thomsen}(2004)}]{10.1098/rsta.2004.1454}%
  \BibitemOpen
  \bibfield  {author} {\bibinfo {author} {\bibfnamefont {S.}~\bibnamefont
  {Reich}}\ and\ \bibinfo {author} {\bibfnamefont {C.}~\bibnamefont
  {Thomsen}},\ }\href {\doibase 10.1098/rsta.2004.1454} {\bibfield  {journal}
  {\bibinfo  {journal} {Phil. Trans. R. Soc. Lond. A}\ }\textbf {\bibinfo
  {volume} {362}},\ \bibinfo {pages} {2271} (\bibinfo {year}
  {2004})}\BibitemShut {NoStop}%
\bibitem [{\citenamefont {Tuinstra}\ and\ \citenamefont
  {Koenig}(1970)}]{10.1063/1.1674108}%
  \BibitemOpen
  \bibfield  {author} {\bibinfo {author} {\bibfnamefont {F.}~\bibnamefont
  {Tuinstra}}\ and\ \bibinfo {author} {\bibfnamefont {J.~L.}\ \bibnamefont
  {Koenig}},\ }\href {\doibase 10.1063/1.1674108} {\bibfield  {journal}
  {\bibinfo  {journal} {J. Chem. Phys.}\ }\textbf {\bibinfo {volume} {53}},\
  \bibinfo {pages} {1126} (\bibinfo {year} {1970})}\BibitemShut {NoStop}%
\bibitem [{\citenamefont {Thomsen}\ and\ \citenamefont
  {Reich}(2000)}]{PhysRevLett.85.5214}%
  \BibitemOpen
  \bibfield  {author} {\bibinfo {author} {\bibfnamefont {C.}~\bibnamefont
  {Thomsen}}\ and\ \bibinfo {author} {\bibfnamefont {S.}~\bibnamefont
  {Reich}},\ }\href@noop {} {\bibfield  {journal} {\bibinfo  {journal} {Phys.
  Rev. Lett.}\ }\textbf {\bibinfo {volume} {85}},\ \bibinfo {pages} {5214}
  (\bibinfo {year} {2000})}\BibitemShut {NoStop}%
\bibitem [{\citenamefont {Maultzsch}\ \emph {et~al.}(2004)\citenamefont
  {Maultzsch}, \citenamefont {Reich},\ and\ \citenamefont
  {Thomsen}}]{PhysRevB.70.155403}%
  \BibitemOpen
  \bibfield  {author} {\bibinfo {author} {\bibfnamefont {J.}~\bibnamefont
  {Maultzsch}}, \bibinfo {author} {\bibfnamefont {S.}~\bibnamefont {Reich}}, \
  and\ \bibinfo {author} {\bibfnamefont {C.}~\bibnamefont {Thomsen}},\ }\href
  {\doibase 10.1103/PhysRevB.70.155403} {\bibfield  {journal} {\bibinfo
  {journal} {Phys. Rev. B}\ }\textbf {\bibinfo {volume} {70}},\ \bibinfo
  {pages} {155403} (\bibinfo {year} {2004})}\BibitemShut {NoStop}%
\bibitem [{\citenamefont {Maultzsch}\ \emph {et~al.}(2001)\citenamefont
  {Maultzsch}, \citenamefont {Reich},\ and\ \citenamefont
  {Thomsen}}]{PhysRevB.64.121407}%
  \BibitemOpen
  \bibfield  {author} {\bibinfo {author} {\bibfnamefont {J.}~\bibnamefont
  {Maultzsch}}, \bibinfo {author} {\bibfnamefont {S.}~\bibnamefont {Reich}}, \
  and\ \bibinfo {author} {\bibfnamefont {C.}~\bibnamefont {Thomsen}},\ }\href
  {\doibase 10.1103/PhysRevB.64.121407} {\bibfield  {journal} {\bibinfo
  {journal} {Phys. Rev. B}\ }\textbf {\bibinfo {volume} {64}},\ \bibinfo
  {pages} {121407} (\bibinfo {year} {2001})}\BibitemShut {NoStop}%
\bibitem [{\citenamefont {Damnjanovi\'{c}}\ \emph {et~al.}(1999)\citenamefont
  {Damnjanovi\'{c}}, \citenamefont {Milo\v{s}evi\'{c}}, \citenamefont
  {Vukovi\'{c}},\ and\ \citenamefont {Sredanovi\'{c}}}]{PhysRevB.60.2728}%
  \BibitemOpen
  \bibfield  {author} {\bibinfo {author} {\bibfnamefont {M.}~\bibnamefont
  {Damnjanovi\'{c}}}, \bibinfo {author} {\bibfnamefont {I.}~\bibnamefont
  {Milo\v{s}evi\'{c}}}, \bibinfo {author} {\bibfnamefont {T.}~\bibnamefont
  {Vukovi\'{c}}}, \ and\ \bibinfo {author} {\bibfnamefont {R.}~\bibnamefont
  {Sredanovi\'{c}}},\ }\href {\doibase 10.1103/PhysRevB.60.2728} {\bibfield
  {journal} {\bibinfo  {journal} {Phys. Rev. B}\ }\textbf {\bibinfo {volume}
  {60}},\ \bibinfo {pages} {2728} (\bibinfo {year} {1999})}\BibitemShut
  {NoStop}%
\bibitem [{\citenamefont {Laudenbach}\ \emph {et~al.}(2013)\citenamefont
  {Laudenbach}, \citenamefont {Hennrich}, \citenamefont {Telg}, \citenamefont
  {Kappes},\ and\ \citenamefont {Maultzsch}}]{PhysRevB.87.165423}%
  \BibitemOpen
  \bibfield  {author} {\bibinfo {author} {\bibfnamefont {J.}~\bibnamefont
  {Laudenbach}}, \bibinfo {author} {\bibfnamefont {F.}~\bibnamefont
  {Hennrich}}, \bibinfo {author} {\bibfnamefont {H.}~\bibnamefont {Telg}},
  \bibinfo {author} {\bibfnamefont {M.}~\bibnamefont {Kappes}}, \ and\ \bibinfo
  {author} {\bibfnamefont {J.}~\bibnamefont {Maultzsch}},\ }\href {\doibase
  10.1103/PhysRevB.87.165423} {\bibfield  {journal} {\bibinfo  {journal} {Phys.
  Rev. B}\ }\textbf {\bibinfo {volume} {87}},\ \bibinfo {pages} {165423}
  (\bibinfo {year} {2013})}\BibitemShut {NoStop}%
\bibitem [{\citenamefont {Souza~Filho}\ \emph {et~al.}(2001)\citenamefont
  {Souza~Filho}, \citenamefont {Jorio}, \citenamefont {Dresselhaus},
  \citenamefont {Dresselhaus}, \citenamefont {Saito}, \citenamefont {Swan},
  \citenamefont {\"Unl\"u}, \citenamefont {Goldberg}, \citenamefont {Hafner},
  \citenamefont {Lieber},\ and\ \citenamefont {Pimenta}}]{PhysRevB.65.035404}%
  \BibitemOpen
  \bibfield  {author} {\bibinfo {author} {\bibfnamefont {A.~G.}\ \bibnamefont
  {Souza~Filho}}, \bibinfo {author} {\bibfnamefont {A.}~\bibnamefont {Jorio}},
  \bibinfo {author} {\bibfnamefont {G.}~\bibnamefont {Dresselhaus}}, \bibinfo
  {author} {\bibfnamefont {M.~S.}\ \bibnamefont {Dresselhaus}}, \bibinfo
  {author} {\bibfnamefont {R.}~\bibnamefont {Saito}}, \bibinfo {author}
  {\bibfnamefont {A.~K.}\ \bibnamefont {Swan}}, \bibinfo {author}
  {\bibfnamefont {M.~S.}\ \bibnamefont {\"Unl\"u}}, \bibinfo {author}
  {\bibfnamefont {B.~B.}\ \bibnamefont {Goldberg}}, \bibinfo {author}
  {\bibfnamefont {J.~H.}\ \bibnamefont {Hafner}}, \bibinfo {author}
  {\bibfnamefont {C.~M.}\ \bibnamefont {Lieber}}, \ and\ \bibinfo {author}
  {\bibfnamefont {M.~A.}\ \bibnamefont {Pimenta}},\ }\href {\doibase
  10.1103/PhysRevB.65.035404} {\bibfield  {journal} {\bibinfo  {journal} {Phys.
  Rev. B}\ }\textbf {\bibinfo {volume} {65}},\ \bibinfo {pages} {035404}
  (\bibinfo {year} {2001})}\BibitemShut {NoStop}%
\bibitem [{\citenamefont {Pimenta}\ \emph {et~al.}(2001)\citenamefont
  {Pimenta}, \citenamefont {Jorio}, \citenamefont {Brown}, \citenamefont
  {Souza~Filho}, \citenamefont {Dresselhaus}, \citenamefont {Hafner},
  \citenamefont {Lieber}, \citenamefont {Saito},\ and\ \citenamefont
  {Dresselhaus}}]{PhysRevB.64.041401}%
  \BibitemOpen
  \bibfield  {author} {\bibinfo {author} {\bibfnamefont {M.~A.}\ \bibnamefont
  {Pimenta}}, \bibinfo {author} {\bibfnamefont {A.}~\bibnamefont {Jorio}},
  \bibinfo {author} {\bibfnamefont {S.~D.~M.}\ \bibnamefont {Brown}}, \bibinfo
  {author} {\bibfnamefont {A.~G.}\ \bibnamefont {Souza~Filho}}, \bibinfo
  {author} {\bibfnamefont {G.}~\bibnamefont {Dresselhaus}}, \bibinfo {author}
  {\bibfnamefont {J.~H.}\ \bibnamefont {Hafner}}, \bibinfo {author}
  {\bibfnamefont {C.~M.}\ \bibnamefont {Lieber}}, \bibinfo {author}
  {\bibfnamefont {R.}~\bibnamefont {Saito}}, \ and\ \bibinfo {author}
  {\bibfnamefont {M.~S.}\ \bibnamefont {Dresselhaus}},\ }\href {\doibase
  10.1103/PhysRevB.64.041401} {\bibfield  {journal} {\bibinfo  {journal} {Phys.
  Rev. B}\ }\textbf {\bibinfo {volume} {64}},\ \bibinfo {pages} {041401}
  (\bibinfo {year} {2001})}\BibitemShut {NoStop}%
\bibitem [{\citenamefont {Souza~Filho}\ \emph {et~al.}(2003)\citenamefont
  {Souza~Filho}, \citenamefont {Jorio}, \citenamefont {Samsonidze},
  \citenamefont {Dresselhaus}, \citenamefont {Pimenta}, \citenamefont
  {Dresselhaus}, \citenamefont {Swan}, \citenamefont {\"Unl\"u}, \citenamefont
  {Goldberg},\ and\ \citenamefont {Saito}}]{PhysRevB.67.035427}%
  \BibitemOpen
  \bibfield  {author} {\bibinfo {author} {\bibfnamefont {A.~G.}\ \bibnamefont
  {Souza~Filho}}, \bibinfo {author} {\bibfnamefont {A.}~\bibnamefont {Jorio}},
  \bibinfo {author} {\bibfnamefont {G.~G.}\ \bibnamefont {Samsonidze}},
  \bibinfo {author} {\bibfnamefont {G.}~\bibnamefont {Dresselhaus}}, \bibinfo
  {author} {\bibfnamefont {M.~A.}\ \bibnamefont {Pimenta}}, \bibinfo {author}
  {\bibfnamefont {M.~S.}\ \bibnamefont {Dresselhaus}}, \bibinfo {author}
  {\bibfnamefont {A.~K.}\ \bibnamefont {Swan}}, \bibinfo {author}
  {\bibfnamefont {M.~S.}\ \bibnamefont {\"Unl\"u}}, \bibinfo {author}
  {\bibfnamefont {B.~B.}\ \bibnamefont {Goldberg}}, \ and\ \bibinfo {author}
  {\bibfnamefont {R.}~\bibnamefont {Saito}},\ }\href {\doibase
  10.1103/PhysRevB.67.035427} {\bibfield  {journal} {\bibinfo  {journal} {Phys.
  Rev. B}\ }\textbf {\bibinfo {volume} {67}},\ \bibinfo {pages} {035427}
  (\bibinfo {year} {2003})}\BibitemShut {NoStop}%
\bibitem [{\citenamefont {Damnjanovi\'{c}}\ \emph {et~al.}(2005)\citenamefont
  {Damnjanovi\'{c}}, \citenamefont {Milo\v{s}evi\'{c}}, \citenamefont
  {Dobard\v{z}i\'{c}}, \citenamefont {Vukovi\'{c}},\ and\ \citenamefont
  {Nikoli\'{c}}}]{damnjanovic2005}%
  \BibitemOpen
  \bibfield  {author} {\bibinfo {author} {\bibfnamefont {M.}~\bibnamefont
  {Damnjanovi\'{c}}}, \bibinfo {author} {\bibfnamefont {I.}~\bibnamefont
  {Milo\v{s}evi\'{c}}}, \bibinfo {author} {\bibfnamefont {E.}~\bibnamefont
  {Dobard\v{z}i\'{c}}}, \bibinfo {author} {\bibfnamefont {T.}~\bibnamefont
  {Vukovi\'{c}}}, \ and\ \bibinfo {author} {\bibfnamefont {B.}~\bibnamefont
  {Nikoli\'{c}}},\ }\enquote {\bibinfo {title} {{A}pplied {P}hysics of
  {N}anotubes; {F}undamentals of {T}heory, {O}ptics and {T}ransport
  {D}evices},}\ \ (\bibinfo  {publisher} {Springer, Berlin},\ \bibinfo {year}
  {2005})\ Chap.~\bibinfo {chapter} {2}\BibitemShut {NoStop}%
\bibitem [{\citenamefont {Gr\"uneis}\ \emph {et~al.}(2009)\citenamefont
  {Gr\"uneis}, \citenamefont {Serrano}, \citenamefont {Bosak}, \citenamefont
  {Lazzeri}, \citenamefont {Molodtsov}, \citenamefont {Wirtz}, \citenamefont
  {Attaccalite}, \citenamefont {Krisch}, \citenamefont {Rubio}, \citenamefont
  {Mauri},\ and\ \citenamefont {Pichler}}]{PhysRevB.80.085423}%
  \BibitemOpen
  \bibfield  {author} {\bibinfo {author} {\bibfnamefont {A.}~\bibnamefont
  {Gr\"uneis}}, \bibinfo {author} {\bibfnamefont {J.}~\bibnamefont {Serrano}},
  \bibinfo {author} {\bibfnamefont {A.}~\bibnamefont {Bosak}}, \bibinfo
  {author} {\bibfnamefont {M.}~\bibnamefont {Lazzeri}}, \bibinfo {author}
  {\bibfnamefont {S.~L.}\ \bibnamefont {Molodtsov}}, \bibinfo {author}
  {\bibfnamefont {L.}~\bibnamefont {Wirtz}}, \bibinfo {author} {\bibfnamefont
  {C.}~\bibnamefont {Attaccalite}}, \bibinfo {author} {\bibfnamefont
  {M.}~\bibnamefont {Krisch}}, \bibinfo {author} {\bibfnamefont
  {A.}~\bibnamefont {Rubio}}, \bibinfo {author} {\bibfnamefont
  {F.}~\bibnamefont {Mauri}}, \ and\ \bibinfo {author} {\bibfnamefont
  {T.}~\bibnamefont {Pichler}},\ }\href {\doibase 10.1103/PhysRevB.80.085423}
  {\bibfield  {journal} {\bibinfo  {journal} {Phys. Rev. B}\ }\textbf {\bibinfo
  {volume} {80}},\ \bibinfo {pages} {085423} (\bibinfo {year}
  {2009})}\BibitemShut {NoStop}%
\bibitem [{\citenamefont {K\"urti}\ \emph {et~al.}(2002)\citenamefont
  {K\"urti}, \citenamefont {Z\'olyomi}, \citenamefont {Gr\"uneis},\ and\
  \citenamefont {Kuzmany}}]{PhysRevB.65.165433}%
  \BibitemOpen
  \bibfield  {author} {\bibinfo {author} {\bibfnamefont {J.}~\bibnamefont
  {K\"urti}}, \bibinfo {author} {\bibfnamefont {V.}~\bibnamefont {Z\'olyomi}},
  \bibinfo {author} {\bibfnamefont {A.}~\bibnamefont {Gr\"uneis}}, \ and\
  \bibinfo {author} {\bibfnamefont {H.}~\bibnamefont {Kuzmany}},\ }\href
  {\doibase 10.1103/PhysRevB.65.165433} {\bibfield  {journal} {\bibinfo
  {journal} {Phys. Rev. B}\ }\textbf {\bibinfo {volume} {65}},\ \bibinfo
  {pages} {165433} (\bibinfo {year} {2002})}\BibitemShut {NoStop}%
\bibitem [{\citenamefont {Laudenbach}\ \emph {et~al.}(2012)\citenamefont
  {Laudenbach}, \citenamefont {Hennrich}, \citenamefont {Kappes},\ and\
  \citenamefont {Maultzsch}}]{10.1002/pssb.201200175}%
  \BibitemOpen
  \bibfield  {author} {\bibinfo {author} {\bibfnamefont {J.}~\bibnamefont
  {Laudenbach}}, \bibinfo {author} {\bibfnamefont {F.}~\bibnamefont
  {Hennrich}}, \bibinfo {author} {\bibfnamefont {M.}~\bibnamefont {Kappes}}, \
  and\ \bibinfo {author} {\bibfnamefont {J.}~\bibnamefont {Maultzsch}},\ }\href
  {\doibase 10.1002/pssb.201200175} {\bibfield  {journal} {\bibinfo  {journal}
  {physica status solidi (b)}\ }\textbf {\bibinfo {volume} {249}},\ \bibinfo
  {pages} {2460} (\bibinfo {year} {2012})}\BibitemShut {NoStop}%
\bibitem [{\citenamefont {Venezuela}\ \emph {et~al.}(2011)\citenamefont
  {Venezuela}, \citenamefont {Lazzeri},\ and\ \citenamefont
  {Mauri}}]{PhysRevB.84.035433}%
  \BibitemOpen
  \bibfield  {author} {\bibinfo {author} {\bibfnamefont {P.}~\bibnamefont
  {Venezuela}}, \bibinfo {author} {\bibfnamefont {M.}~\bibnamefont {Lazzeri}},
  \ and\ \bibinfo {author} {\bibfnamefont {F.}~\bibnamefont {Mauri}},\ }\href
  {\doibase 10.1103/PhysRevB.84.035433} {\bibfield  {journal} {\bibinfo
  {journal} {Phys. Rev. B}\ }\textbf {\bibinfo {volume} {84}},\ \bibinfo
  {pages} {035433} (\bibinfo {year} {2011})}\BibitemShut {NoStop}%
\bibitem [{\citenamefont {Damnjanovi\ifmmode~\acute{c}\else \'{c}\fi{}}\ \emph
  {et~al.}(2000)\citenamefont {Damnjanovi\ifmmode~\acute{c}\else \'{c}\fi{}},
  \citenamefont {Vukovi\ifmmode~\acute{c}\else \'{c}\fi{}},\ and\ \citenamefont
  {Milo\ifmmode \check{s}\else \v{s}\fi{}evi\ifmmode~\acute{c}\else
  \'{c}\fi{}}}]{10.1088/0305-4470/33/37/308}%
  \BibitemOpen
  \bibfield  {author} {\bibinfo {author} {\bibfnamefont {M.}~\bibnamefont
  {Damnjanovi\ifmmode~\acute{c}\else \'{c}\fi{}}}, \bibinfo {author}
  {\bibfnamefont {T.}~\bibnamefont {Vukovi\ifmmode~\acute{c}\else \'{c}\fi{}}},
  \ and\ \bibinfo {author} {\bibfnamefont {I.}~\bibnamefont {Milo\ifmmode
  \check{s}\else \v{s}\fi{}evi\ifmmode~\acute{c}\else \'{c}\fi{}}},\ }\href
  {\doibase 10.1088/0305-4470/33/37/308} {\bibfield  {journal} {\bibinfo
  {journal} {J. Phys. A: Math. Gen.}\ }\textbf {\bibinfo {volume} {33}},\
  \bibinfo {pages} {6561 } (\bibinfo {year} {2000})}\BibitemShut {NoStop}%
\bibitem [{\citenamefont {Maultzsch}\ \emph {et~al.}(2002)\citenamefont
  {Maultzsch}, \citenamefont {Reich}, \citenamefont {Thomsen}, \citenamefont
  {Dobard\v{z}i\'{c}}, \citenamefont {Milo\v{s}evi\'{c}},\ and\ \citenamefont
  {Damnjanovi\'{c}}}]{10.1016/S0038-1098(02)00025-X}%
  \BibitemOpen
  \bibfield  {author} {\bibinfo {author} {\bibfnamefont {J.}~\bibnamefont
  {Maultzsch}}, \bibinfo {author} {\bibfnamefont {S.}~\bibnamefont {Reich}},
  \bibinfo {author} {\bibfnamefont {C.}~\bibnamefont {Thomsen}}, \bibinfo
  {author} {\bibfnamefont {E.}~\bibnamefont {Dobard\v{z}i\'{c}}}, \bibinfo
  {author} {\bibfnamefont {I.}~\bibnamefont {Milo\v{s}evi\'{c}}}, \ and\
  \bibinfo {author} {\bibfnamefont {M.}~\bibnamefont {Damnjanovi\'{c}}},\
  }\href {\doibase http://dx.doi.org/10.1016/S0038-1098(02)00025-X} {\bibfield
  {journal} {\bibinfo  {journal} {Solid State Communications}\ }\textbf
  {\bibinfo {volume} {121}},\ \bibinfo {pages} {471 } (\bibinfo {year}
  {2002})}\BibitemShut {NoStop}%
\bibitem [{Note1()}]{Note1}%
  \BibitemOpen
  \bibinfo {note} {See the Supplemental Material at http: which includes Refs.
  \protect \rev@citealpnum {10.1088/0305-4470/33/37/308,
  10.1016/S0038-1098(02)00025-X, reich2004, 10.1016/0009-2614(93)87205-H,
  PhysRevLett.93.177401, PhysRevB.72.205438, PhysRevB.74.115415,
  PhysRevB.71.035416, 10.1016/j.carbon.2011.06.076, PhysRevB.74.184306,
  PhysRevB.59.8355,
  10.1021/nn2044356,PhysRevB.60.2728,LaudenbachDissertation}.}\BibitemShut
  {Stop}%
\bibitem [{\citenamefont {Popov}\ and\ \citenamefont
  {Henrard}(2004)}]{PhysRevB.70.115407}%
  \BibitemOpen
  \bibfield  {author} {\bibinfo {author} {\bibfnamefont {V.~N.}\ \bibnamefont
  {Popov}}\ and\ \bibinfo {author} {\bibfnamefont {L.}~\bibnamefont
  {Henrard}},\ }\href {\doibase 10.1103/PhysRevB.70.115407} {\bibfield
  {journal} {\bibinfo  {journal} {Phys. Rev. B}\ }\textbf {\bibinfo {volume}
  {70}},\ \bibinfo {pages} {115407} (\bibinfo {year} {2004})}\BibitemShut
  {NoStop}%
\bibitem [{\citenamefont {G\"ulseren}\ \emph {et~al.}(2002)\citenamefont
  {G\"ulseren}, \citenamefont {Yildirim},\ and\ \citenamefont
  {Ciraci}}]{PhysRevB.65.153405}%
  \BibitemOpen
  \bibfield  {author} {\bibinfo {author} {\bibfnamefont {O.}~\bibnamefont
  {G\"ulseren}}, \bibinfo {author} {\bibfnamefont {T.}~\bibnamefont
  {Yildirim}}, \ and\ \bibinfo {author} {\bibfnamefont {S.}~\bibnamefont
  {Ciraci}},\ }\href {\doibase 10.1103/PhysRevB.65.153405} {\bibfield
  {journal} {\bibinfo  {journal} {Phys. Rev. B}\ }\textbf {\bibinfo {volume}
  {65}},\ \bibinfo {pages} {153405} (\bibinfo {year} {2002})}\BibitemShut
  {NoStop}%
\bibitem [{\citenamefont {Popov}(2004)}]{10.1088/1367-2630/6/1/017}%
  \BibitemOpen
  \bibfield  {author} {\bibinfo {author} {\bibfnamefont {V.~N.}\ \bibnamefont
  {Popov}},\ }\href {\doibase 10.1088/1367-2630/6/1/017} {\bibfield  {journal}
  {\bibinfo  {journal} {New Journal of Physics}\ }\textbf {\bibinfo {volume}
  {6}},\ \bibinfo {pages} {17} (\bibinfo {year} {2004})}\BibitemShut {NoStop}%
\bibitem [{\citenamefont {Kataura}\ \emph {et~al.}(1999)\citenamefont
  {Kataura}, \citenamefont {Kumazawa}, \citenamefont {Maniwa}, \citenamefont
  {Umezu}, \citenamefont {Suzuki}, \citenamefont {Ohtsuka},\ and\ \citenamefont
  {Achiba}}]{10.1016/S0379-6779(98)00278-1}%
  \BibitemOpen
  \bibfield  {author} {\bibinfo {author} {\bibfnamefont {H.}~\bibnamefont
  {Kataura}}, \bibinfo {author} {\bibfnamefont {Y.}~\bibnamefont {Kumazawa}},
  \bibinfo {author} {\bibfnamefont {Y.}~\bibnamefont {Maniwa}}, \bibinfo
  {author} {\bibfnamefont {I.}~\bibnamefont {Umezu}}, \bibinfo {author}
  {\bibfnamefont {S.}~\bibnamefont {Suzuki}}, \bibinfo {author} {\bibfnamefont
  {Y.}~\bibnamefont {Ohtsuka}}, \ and\ \bibinfo {author} {\bibfnamefont
  {Y.}~\bibnamefont {Achiba}},\ }\href {\doibase 10.1016/S0379-6779(98)00278-1}
  {\bibfield  {journal} {\bibinfo  {journal} {Synthetic Metals}\ }\textbf
  {\bibinfo {volume} {103}},\ \bibinfo {pages} {2555 } (\bibinfo {year}
  {1999})}\BibitemShut {NoStop}%
\bibitem [{\citenamefont {Laudenbach}(2014)}]{LaudenbachDissertation}%
  \BibitemOpen
  \bibfield  {author} {\bibinfo {author} {\bibfnamefont {J.}~\bibnamefont
  {Laudenbach}},\ }\href@noop {} {Ph.D. thesis},\ \bibinfo  {school} {TU
  Berlin} (\bibinfo {year} {2014}),\ \bibinfo {note}
  {urn:nbn:de:kobv:83-opus4-46019}\BibitemShut {NoStop}%
\bibitem [{\citenamefont {Laudenbach}\ \emph {et~al.}(2015)\citenamefont
  {Laudenbach}, \citenamefont {Schmid}, \citenamefont {Herziger}, \citenamefont
  {Hennrich}, \citenamefont {Kappes}, \citenamefont {Muoth}, \citenamefont
  {Haluska}, \citenamefont {Hof}, \citenamefont {Backes}, \citenamefont
  {Hauke}, \citenamefont {Hirsch},\ and\ \citenamefont
  {Maultzsch}}]{Laudenbach2015}%
  \BibitemOpen
  \bibfield  {author} {\bibinfo {author} {\bibfnamefont {J.}~\bibnamefont
  {Laudenbach}}, \bibinfo {author} {\bibfnamefont {D.}~\bibnamefont {Schmid}},
  \bibinfo {author} {\bibfnamefont {F.}~\bibnamefont {Herziger}}, \bibinfo
  {author} {\bibfnamefont {F.}~\bibnamefont {Hennrich}}, \bibinfo {author}
  {\bibfnamefont {M.}~\bibnamefont {Kappes}}, \bibinfo {author} {\bibfnamefont
  {M.}~\bibnamefont {Muoth}}, \bibinfo {author} {\bibfnamefont
  {M.}~\bibnamefont {Haluska}}, \bibinfo {author} {\bibfnamefont
  {F.}~\bibnamefont {Hof}}, \bibinfo {author} {\bibfnamefont {C.}~\bibnamefont
  {Backes}}, \bibinfo {author} {\bibfnamefont {F.}~\bibnamefont {Hauke}},
  \bibinfo {author} {\bibfnamefont {A.}~\bibnamefont {Hirsch}}, \ and\ \bibinfo
  {author} {\bibfnamefont {J.}~\bibnamefont {Maultzsch}},\ }\href@noop {}
  {\bibfield  {journal} {\bibinfo  {journal} {submitted}\ } (\bibinfo {year}
  {2015})}\BibitemShut {NoStop}%
\bibitem [{\citenamefont {Telg}\ \emph {et~al.}(2012)\citenamefont {Telg},
  \citenamefont {Duque}, \citenamefont {Staiger}, \citenamefont {Tu},
  \citenamefont {Hennrich}, \citenamefont {Kappes}, \citenamefont {Zheng},
  \citenamefont {Maultzsch}, \citenamefont {Thomsen},\ and\ \citenamefont
  {Doorn}}]{10.1021/nn2044356}%
  \BibitemOpen
  \bibfield  {author} {\bibinfo {author} {\bibfnamefont {H.}~\bibnamefont
  {Telg}}, \bibinfo {author} {\bibfnamefont {J.~G.}\ \bibnamefont {Duque}},
  \bibinfo {author} {\bibfnamefont {M.}~\bibnamefont {Staiger}}, \bibinfo
  {author} {\bibfnamefont {X.}~\bibnamefont {Tu}}, \bibinfo {author}
  {\bibfnamefont {F.}~\bibnamefont {Hennrich}}, \bibinfo {author}
  {\bibfnamefont {M.~M.}\ \bibnamefont {Kappes}}, \bibinfo {author}
  {\bibfnamefont {M.}~\bibnamefont {Zheng}}, \bibinfo {author} {\bibfnamefont
  {J.}~\bibnamefont {Maultzsch}}, \bibinfo {author} {\bibfnamefont
  {C.}~\bibnamefont {Thomsen}}, \ and\ \bibinfo {author} {\bibfnamefont
  {S.~K.}\ \bibnamefont {Doorn}},\ }\href {\doibase 10.1021/nn2044356}
  {\bibfield  {journal} {\bibinfo  {journal} {ACS Nano}\ }\textbf {\bibinfo
  {volume} {6}},\ \bibinfo {pages} {904} (\bibinfo {year} {2012})}\BibitemShut
  {NoStop}%
\bibitem [{\citenamefont {K\"urti}\ \emph {et~al.}(1998)\citenamefont
  {K\"urti}, \citenamefont {Kresse},\ and\ \citenamefont
  {Kuzmany}}]{PhysRevB.58.R8869}%
  \BibitemOpen
  \bibfield  {author} {\bibinfo {author} {\bibfnamefont {J.}~\bibnamefont
  {K\"urti}}, \bibinfo {author} {\bibfnamefont {G.}~\bibnamefont {Kresse}}, \
  and\ \bibinfo {author} {\bibfnamefont {H.}~\bibnamefont {Kuzmany}},\ }\href
  {\doibase 10.1103/PhysRevB.58.R8869} {\bibfield  {journal} {\bibinfo
  {journal} {Phys. Rev. B}\ }\textbf {\bibinfo {volume} {58}},\ \bibinfo
  {pages} {R8869} (\bibinfo {year} {1998})}\BibitemShut {NoStop}%
\bibitem [{\citenamefont {Popov}\ and\ \citenamefont
  {Lambin}(2006)}]{PhysRevB.73.085407}%
  \BibitemOpen
  \bibfield  {author} {\bibinfo {author} {\bibfnamefont {V.~N.}\ \bibnamefont
  {Popov}}\ and\ \bibinfo {author} {\bibfnamefont {P.}~\bibnamefont {Lambin}},\
  }\href {\doibase 10.1103/PhysRevB.73.085407} {\bibfield  {journal} {\bibinfo
  {journal} {Phys. Rev. B}\ }\textbf {\bibinfo {volume} {73}},\ \bibinfo
  {pages} {085407} (\bibinfo {year} {2006})}\BibitemShut {NoStop}%
\bibitem [{\citenamefont {Telg}\ \emph {et~al.}(2004)\citenamefont {Telg},
  \citenamefont {Maultzsch}, \citenamefont {Reich}, \citenamefont {Hennrich},\
  and\ \citenamefont {Thomsen}}]{PhysRevLett.93.177401}%
  \BibitemOpen
  \bibfield  {author} {\bibinfo {author} {\bibfnamefont {H.}~\bibnamefont
  {Telg}}, \bibinfo {author} {\bibfnamefont {J.}~\bibnamefont {Maultzsch}},
  \bibinfo {author} {\bibfnamefont {S.}~\bibnamefont {Reich}}, \bibinfo
  {author} {\bibfnamefont {F.}~\bibnamefont {Hennrich}}, \ and\ \bibinfo
  {author} {\bibfnamefont {C.}~\bibnamefont {Thomsen}},\ }\href {\doibase
  10.1103/PhysRevLett.93.177401} {\bibfield  {journal} {\bibinfo  {journal}
  {Phys. Rev. Lett.}\ }\textbf {\bibinfo {volume} {93}},\ \bibinfo {pages}
  {177401} (\bibinfo {year} {2004})}\BibitemShut {NoStop}%
\bibitem [{\citenamefont {Ferrari}\ and\ \citenamefont
  {Basko}(2013)}]{10.1038/nnano.2013.46}%
  \BibitemOpen
  \bibfield  {author} {\bibinfo {author} {\bibfnamefont {A.~C.}\ \bibnamefont
  {Ferrari}}\ and\ \bibinfo {author} {\bibfnamefont {D.~M.}\ \bibnamefont
  {Basko}},\ }\href {\doibase 10.1038/nnano.2013.46} {\bibfield  {journal}
  {\bibinfo  {journal} {Nature Nanotech.}\ }\textbf {\bibinfo {volume} {8}},\
  \bibinfo {pages} {235–246} (\bibinfo {year} {2013})}\BibitemShut {NoStop}%
\bibitem [{\citenamefont {Jishi}\ \emph {et~al.}(1993)\citenamefont {Jishi},
  \citenamefont {Venkataraman}, \citenamefont {Dresselhaus},\ and\
  \citenamefont {Dresselhaus}}]{10.1016/0009-2614(93)87205-H}%
  \BibitemOpen
  \bibfield  {author} {\bibinfo {author} {\bibfnamefont {R.}~\bibnamefont
  {Jishi}}, \bibinfo {author} {\bibfnamefont {L.}~\bibnamefont {Venkataraman}},
  \bibinfo {author} {\bibfnamefont {M.}~\bibnamefont {Dresselhaus}}, \ and\
  \bibinfo {author} {\bibfnamefont {G.}~\bibnamefont {Dresselhaus}},\ }\href
  {\doibase 10.1016/0009-2614(93)87205-H} {\bibfield  {journal} {\bibinfo
  {journal} {Chem. Phys. Lett.}\ }\textbf {\bibinfo {volume} {209}},\ \bibinfo
  {pages} {77 } (\bibinfo {year} {1993})}\BibitemShut {NoStop}%
\bibitem [{\citenamefont {Maultzsch}\ \emph {et~al.}(2005)\citenamefont
  {Maultzsch}, \citenamefont {Telg}, \citenamefont {Reich},\ and\ \citenamefont
  {Thomsen}}]{PhysRevB.72.205438}%
  \BibitemOpen
  \bibfield  {author} {\bibinfo {author} {\bibfnamefont {J.}~\bibnamefont
  {Maultzsch}}, \bibinfo {author} {\bibfnamefont {H.}~\bibnamefont {Telg}},
  \bibinfo {author} {\bibfnamefont {S.}~\bibnamefont {Reich}}, \ and\ \bibinfo
  {author} {\bibfnamefont {C.}~\bibnamefont {Thomsen}},\ }\href {\doibase
  10.1103/PhysRevB.72.205438} {\bibfield  {journal} {\bibinfo  {journal} {Phys.
  Rev. B}\ }\textbf {\bibinfo {volume} {72}},\ \bibinfo {pages} {205438}
  (\bibinfo {year} {2005})}\BibitemShut {NoStop}%
\bibitem [{\citenamefont {Telg}\ \emph {et~al.}(2006)\citenamefont {Telg},
  \citenamefont {Maultzsch}, \citenamefont {Reich},\ and\ \citenamefont
  {Thomsen}}]{PhysRevB.74.115415}%
  \BibitemOpen
  \bibfield  {author} {\bibinfo {author} {\bibfnamefont {H.}~\bibnamefont
  {Telg}}, \bibinfo {author} {\bibfnamefont {J.}~\bibnamefont {Maultzsch}},
  \bibinfo {author} {\bibfnamefont {S.}~\bibnamefont {Reich}}, \ and\ \bibinfo
  {author} {\bibfnamefont {C.}~\bibnamefont {Thomsen}},\ }\href {\doibase
  10.1103/PhysRevB.74.115415} {\bibfield  {journal} {\bibinfo  {journal} {Phys.
  Rev. B}\ }\textbf {\bibinfo {volume} {74}},\ \bibinfo {pages} {115415}
  (\bibinfo {year} {2006})}\BibitemShut {NoStop}%
\bibitem [{\citenamefont {Mach\'on}\ \emph {et~al.}(2005)\citenamefont
  {Mach\'on}, \citenamefont {Reich}, \citenamefont {Telg}, \citenamefont
  {Maultzsch}, \citenamefont {Ordej\'on},\ and\ \citenamefont
  {Thomsen}}]{PhysRevB.71.035416}%
  \BibitemOpen
  \bibfield  {author} {\bibinfo {author} {\bibfnamefont {M.}~\bibnamefont
  {Mach\'on}}, \bibinfo {author} {\bibfnamefont {S.}~\bibnamefont {Reich}},
  \bibinfo {author} {\bibfnamefont {H.}~\bibnamefont {Telg}}, \bibinfo {author}
  {\bibfnamefont {J.}~\bibnamefont {Maultzsch}}, \bibinfo {author}
  {\bibfnamefont {P.}~\bibnamefont {Ordej\'on}}, \ and\ \bibinfo {author}
  {\bibfnamefont {C.}~\bibnamefont {Thomsen}},\ }\href {\doibase
  10.1103/PhysRevB.71.035416} {\bibfield  {journal} {\bibinfo  {journal} {Phys.
  Rev. B}\ }\textbf {\bibinfo {volume} {71}},\ \bibinfo {pages} {035416}
  (\bibinfo {year} {2005})}\BibitemShut {NoStop}%
\bibitem [{\citenamefont {Costa}\ \emph {et~al.}(2011)\citenamefont {Costa},
  \citenamefont {Fantini}, \citenamefont {Righi}, \citenamefont {Bachmatiuk},
  \citenamefont {Rümmeli}, \citenamefont {Saito},\ and\ \citenamefont
  {Pimenta}}]{10.1016/j.carbon.2011.06.076}%
  \BibitemOpen
  \bibfield  {author} {\bibinfo {author} {\bibfnamefont {S.~D.}\ \bibnamefont
  {Costa}}, \bibinfo {author} {\bibfnamefont {C.}~\bibnamefont {Fantini}},
  \bibinfo {author} {\bibfnamefont {A.}~\bibnamefont {Righi}}, \bibinfo
  {author} {\bibfnamefont {A.}~\bibnamefont {Bachmatiuk}}, \bibinfo {author}
  {\bibfnamefont {M.~H.}\ \bibnamefont {Rümmeli}}, \bibinfo {author}
  {\bibfnamefont {R.}~\bibnamefont {Saito}}, \ and\ \bibinfo {author}
  {\bibfnamefont {M.~A.}\ \bibnamefont {Pimenta}},\ }\href {\doibase
  10.1016/j.carbon.2011.06.076} {\bibfield  {journal} {\bibinfo  {journal}
  {Carbon}\ }\textbf {\bibinfo {volume} {49}},\ \bibinfo {pages} {4719 }
  (\bibinfo {year} {2011})}\BibitemShut {NoStop}%
\bibitem [{\citenamefont {Di~Donato}\ \emph {et~al.}(2006)\citenamefont
  {Di~Donato}, \citenamefont {Tommasini}, \citenamefont {Castiglioni},\ and\
  \citenamefont {Zerbi}}]{PhysRevB.74.184306}%
  \BibitemOpen
  \bibfield  {author} {\bibinfo {author} {\bibfnamefont {E.}~\bibnamefont
  {Di~Donato}}, \bibinfo {author} {\bibfnamefont {M.}~\bibnamefont
  {Tommasini}}, \bibinfo {author} {\bibfnamefont {C.}~\bibnamefont
  {Castiglioni}}, \ and\ \bibinfo {author} {\bibfnamefont {G.}~\bibnamefont
  {Zerbi}},\ }\href {\doibase 10.1103/PhysRevB.74.184306} {\bibfield  {journal}
  {\bibinfo  {journal} {Phys. Rev. B}\ }\textbf {\bibinfo {volume} {74}},\
  \bibinfo {pages} {184306} (\bibinfo {year} {2006})}\BibitemShut {NoStop}%
\bibitem [{\citenamefont {Popov}\ \emph {et~al.}(1999)\citenamefont {Popov},
  \citenamefont {Van~Doren},\ and\ \citenamefont
  {Balkanski}}]{PhysRevB.59.8355}%
  \BibitemOpen
  \bibfield  {author} {\bibinfo {author} {\bibfnamefont {V.~N.}\ \bibnamefont
  {Popov}}, \bibinfo {author} {\bibfnamefont {V.~E.}\ \bibnamefont
  {Van~Doren}}, \ and\ \bibinfo {author} {\bibfnamefont {M.}~\bibnamefont
  {Balkanski}},\ }\href {\doibase 10.1103/PhysRevB.59.8355} {\bibfield
  {journal} {\bibinfo  {journal} {Phys. Rev. B}\ }\textbf {\bibinfo {volume}
  {59}},\ \bibinfo {pages} {8355} (\bibinfo {year} {1999})}\BibitemShut
  {NoStop}%
\end{thebibliography}

\newpage
\includepdf{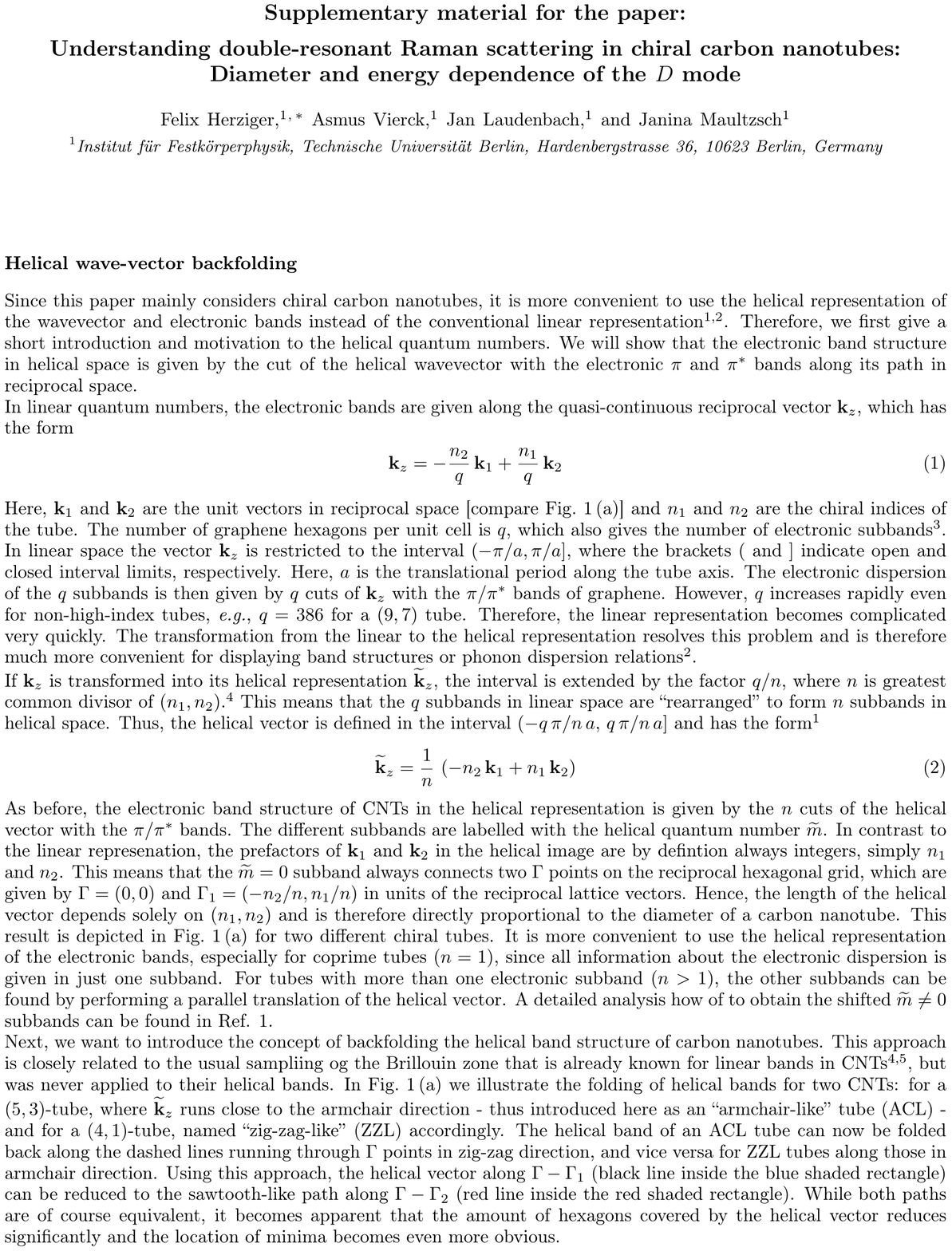}

~
\newpage
\includepdf{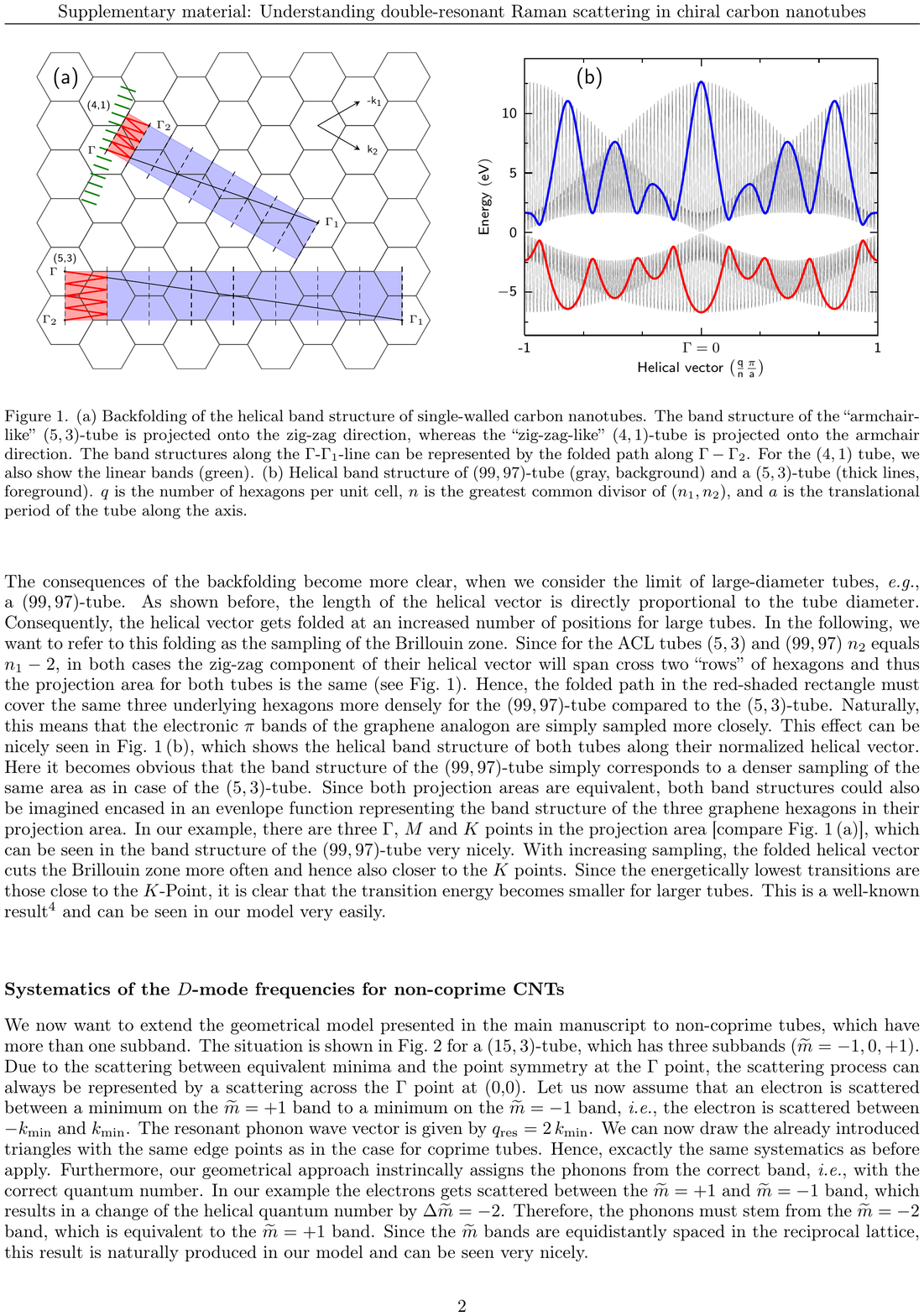}

~
\newpage
\includepdf{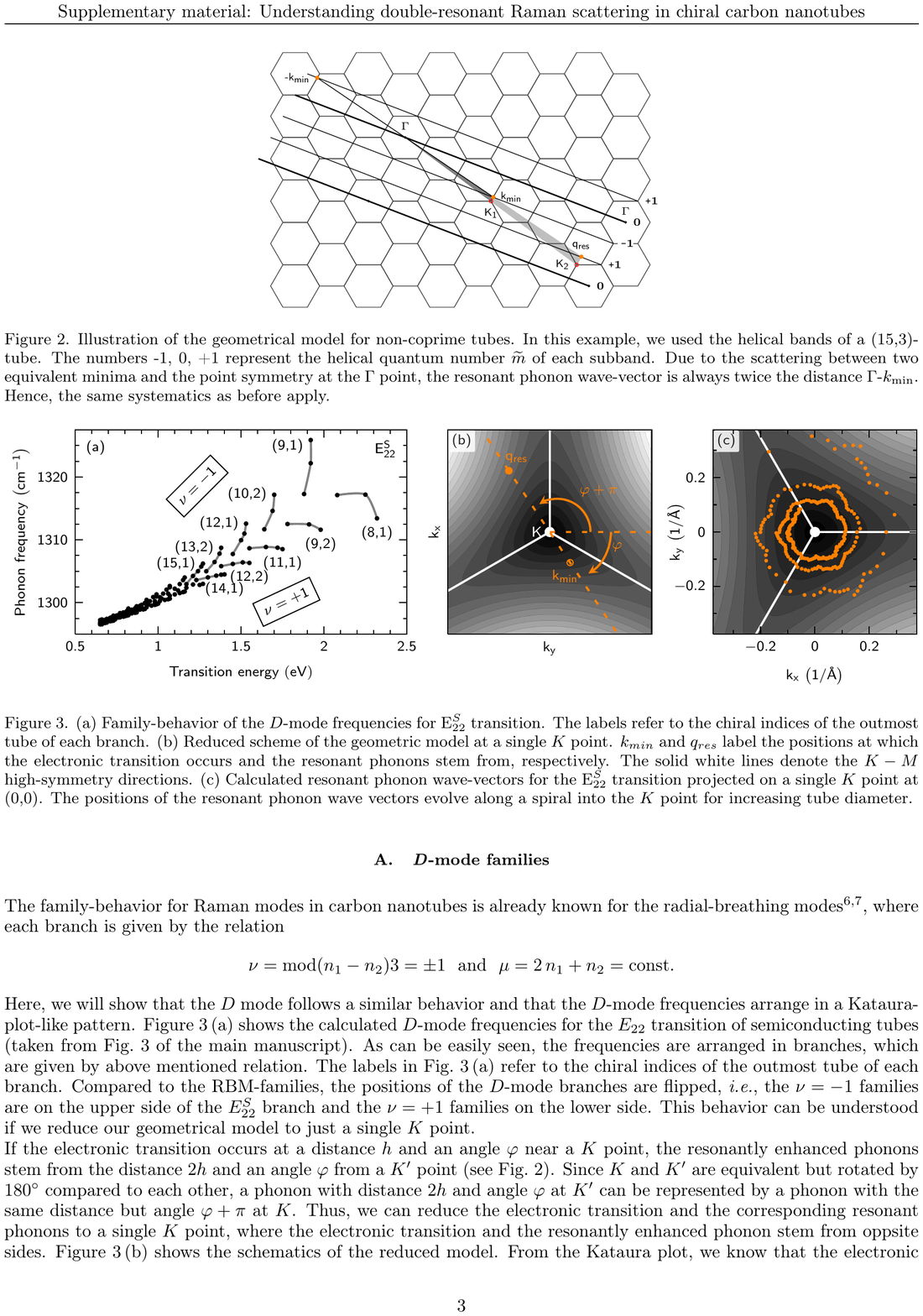}

~
\newpage
\includepdf{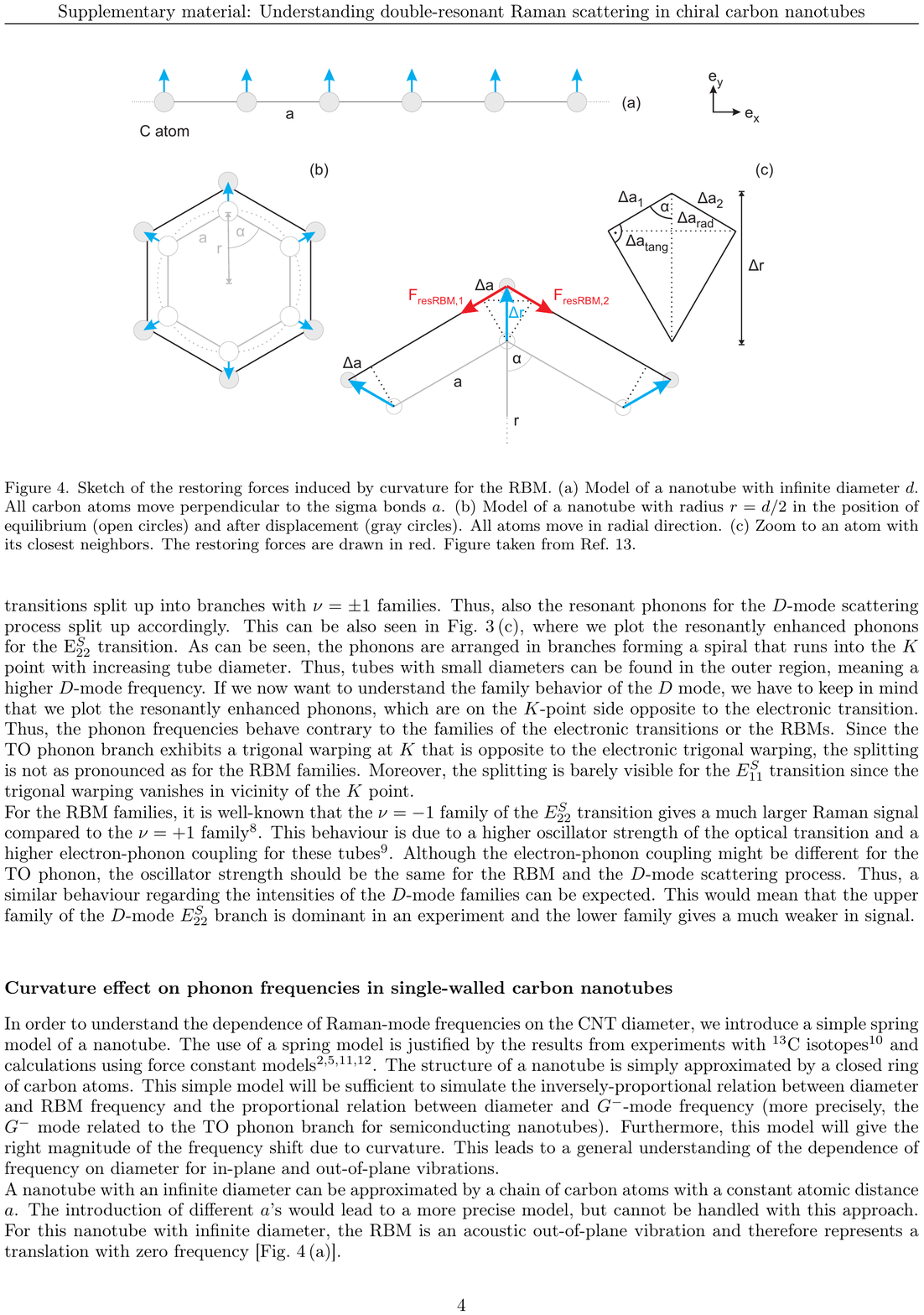}

~
\newpage
\includepdf{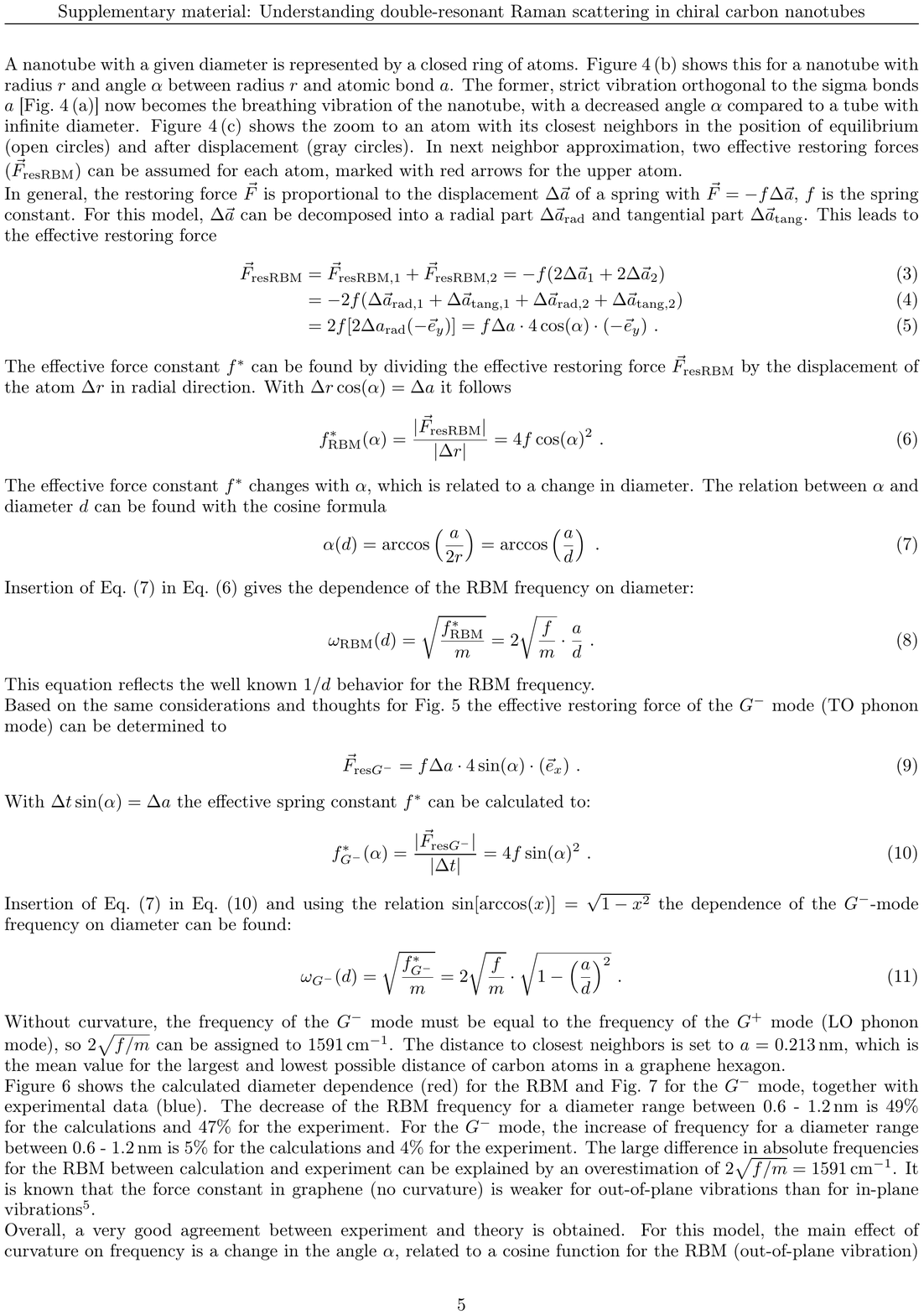}

~
\newpage
\includepdf{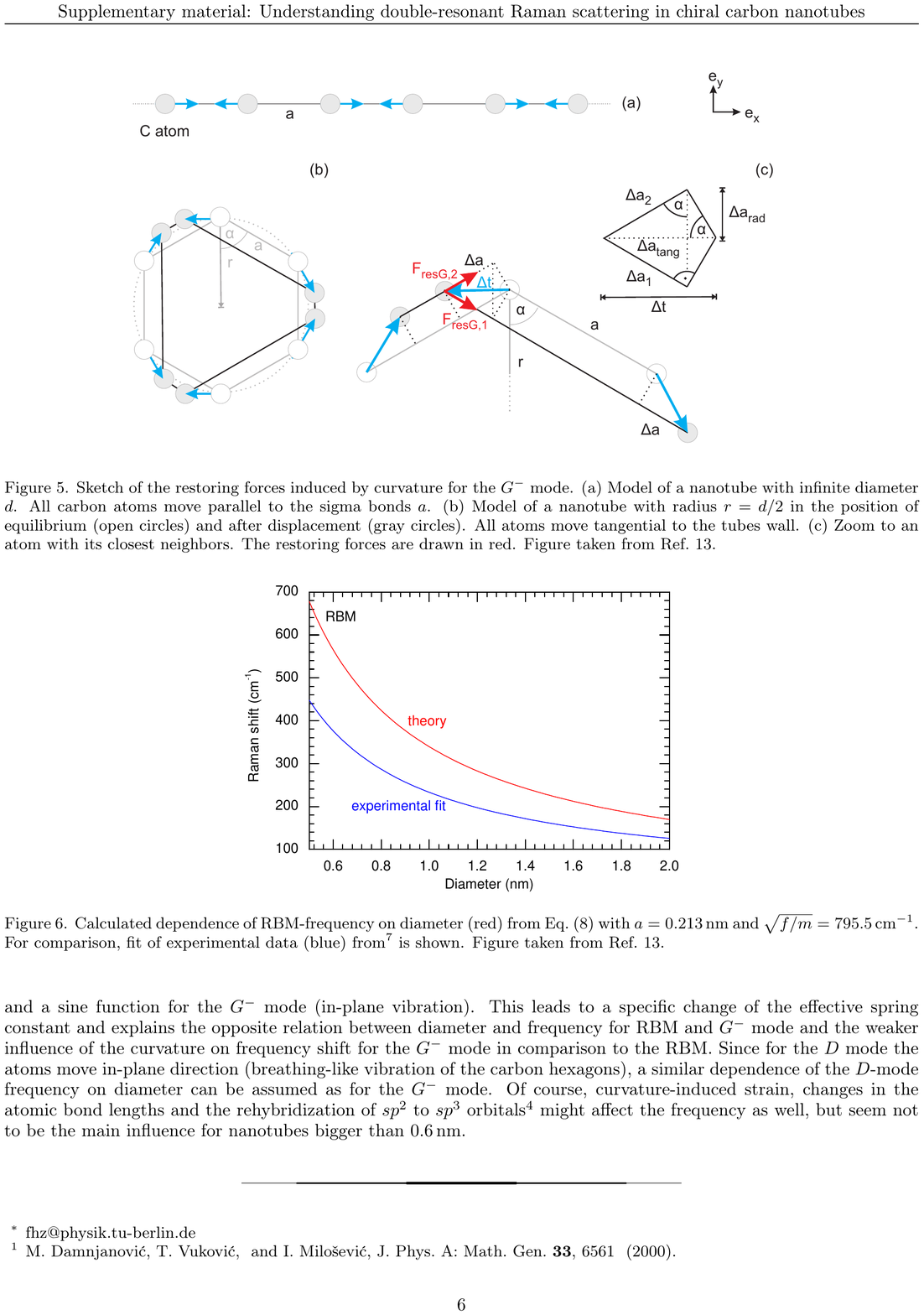}

~
\newpage
\includepdf{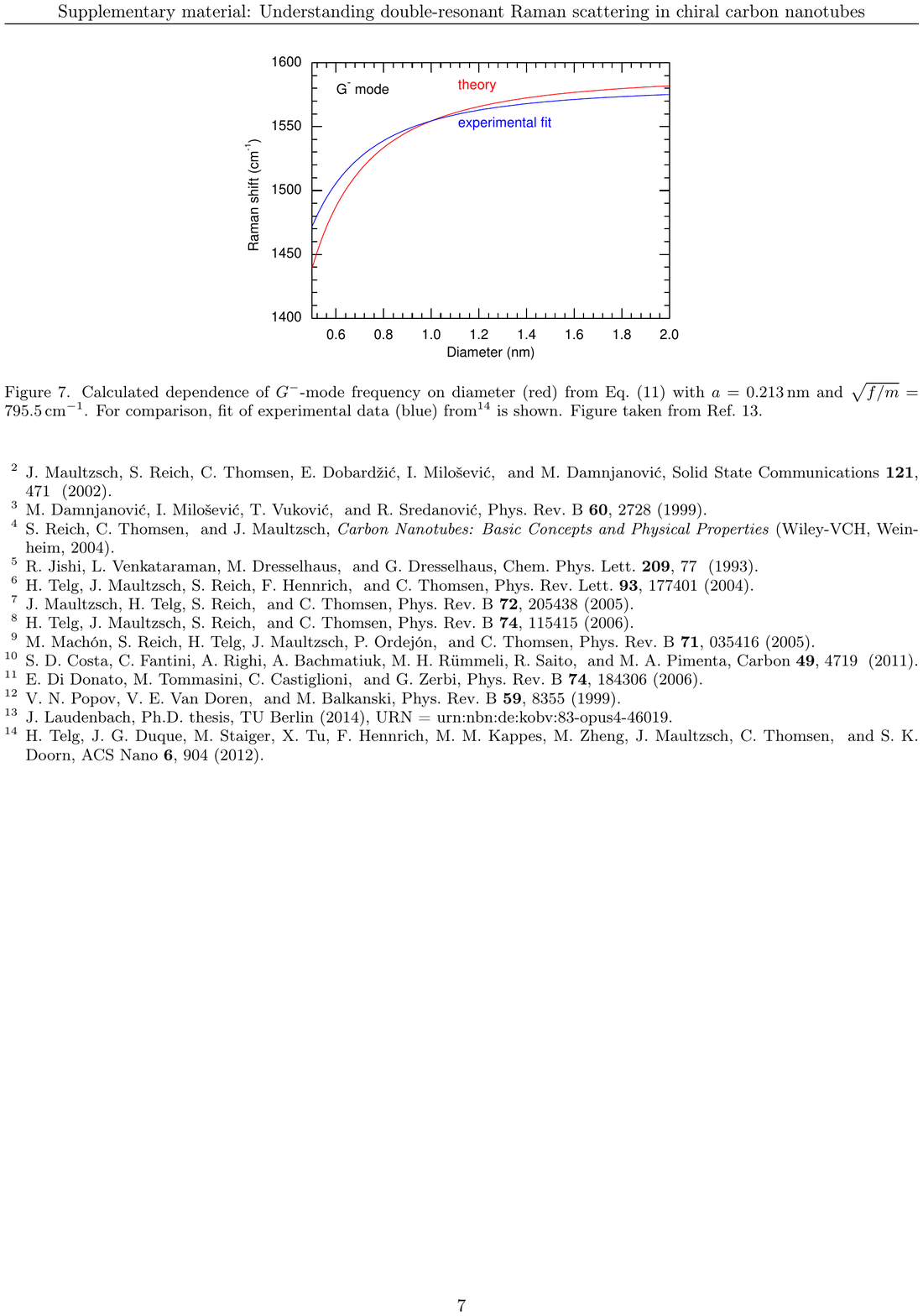}

\end{document}